\documentclass{aastex}
\usepackage{amsmath}
\usepackage{amssymb}
\usepackage[numberedappendix]{emulateapj5}
\usepackage{apjfonts}
\usepackage{epsfig}

\makeatletter

\newenvironment{inlinefigure}{%
\def\@captype{figure}%
\noindent\begin{minipage}{0.999\linewidth}\begin{center}}
{\end{center}\end{minipage}\smallskip}
\makeatother

\newcommand{\kms}{\,{\rm km\,s^{-1}}}

\renewcommand{\day}{\,{\rm d}}
\newcommand{\yr}{\,{\rm yr}}

\newcommand{\gyr}{\,{\rm Gyr}}

\newcommand{\ergs}{\,{\rm ergs\,s^{-1}}}

\newcommand{\msun}{\,M_\odot}

\newcommand{\mdot}{\,M_\odot\,{\rm yr}^{-1}}

\newcommand{\flux}{\,{\rm ergs\,s^{-1}\,cm^{-2}}}

\newcommand{\be}{\begin{equation}}
\newcommand{\ee}{\end{equation}}

\newcommand{\bea}{\begin{eqnarray}}
\newcommand{\eea}{\end{eqnarray}}

\newcommand{\ben}{\begin{enumerate}}
\newcommand{\een}{\end{enumerate}}

\begin{document}

\shorttitle{Low- and Intermediate-Mass X-ray Binaries}
\shortauthors{PFAHL, RAPPAPORT, PODSIADLOWSKI}


\submitted{Submitted to ApJ}

\title{The Galactic Population of Low- and Intermediate-Mass X-ray
Binaries}

\author{Eric Pfahl\altaffilmark{1}, Saul Rappaport\altaffilmark{2}, 
and Philipp Podsiadlowski\altaffilmark{3}}

\altaffiltext{1}{Chandra Fellow; Harvard-Smithsonian Center for 
Astrophysics, 60 Garden Street, Cambridge, MA 02138; epfahl@cfa.harvard.edu} 
\altaffiltext{2}{Center for Space
Research, Massachusetts Institute of Technology, Cambridge, MA, 02139;
sar@mit.edu} 
\altaffiltext{3}{Department of
Astrophysics, Oxford University, Oxford, OX1 3RH, England, UK;
podsi@astro.ox.ac.uk}


\begin{abstract}

We present the first study that combines binary population synthesis
in the Galactic disk and detailed evolutionary calculations of low-
and intermediate-mass X-ray binaries (L/IMXBs).  This approach allows
us to follow completely the formation of L/IMXBs, and their evolution
through the X-ray phase, to the point when they become binary
millisecond pulsars (BMPs).  We show that the formation probability
of IMXBs with initial donor masses of 1.5--$4\msun$ is typically
$\ga$5 times higher than that of standard LMXBs with initial donor
masses of $<$$1.5\msun$.  Since IMXBs evolve to resemble observed
LMXBs, we suggest that the majority of the observed systems may have
descended from IMXBs.  Distributions at the current epoch of the
orbital periods, donor masses, and mass accretion rates of L/IMXBs
have been computed, as have orbital-period distributions of BMPs.
Several significant discrepancies between the theoretical and observed
distributions are discussed.  We find that the total number of luminous
($L_{\rm X} > 10^{36}\ergs$) X-ray sources at the current epoch and
the period distribution of BMPs are very sensitive to the parameters
in analytic formula describing the common-envelope phase that precedes
the formation of the neutron star.  The orbital-period distribution of
observed BMPs strongly favors cases where the common envelope is more
easily ejected.  However, this leads to a $\ga$100-fold overproduction
of the theoretical number of luminous X-ray sources relative to the
total observed number of LMXBs.  X-ray irradiation of the donor star
may result in a dramatic reduction in the X-ray active lifetime of
L/IMXBs, thus possibly resolving the overproduction problem, as well
as the long-standing BMP/LMXB birthrate problem.

\end{abstract}




\section{INTRODUCTION}\label{sec:intro}

Roughly 140 low-mass X-ray binaries (LMXBs) are known in the Galaxy
\citep{liu01}, with orbital periods from 11\,min to $\sim$1\,yr, donor
masses of $\sim$0.01--$2\msun$, and X-ray luminosities from the
detection sensitivities to $\sim$$10^{38}\ergs$.  Over the past twenty
years or so, many theoretical studies of LMXBs have aimed at
accounting for their abundance and variety.  During this time, a
standard picture for the formation and evolution of LMXBs in the
Galactic disk has emerged.  However, recent observational and
theoretical work have challenged the conventional wisdom and prompted
a renewed interest in the origins of observed LMXBs.  Specifically, it
has been realized that many, perhaps even the majority, of the
identified LMXBs with low-mass stellar companions may be descendants
of systems with {\em intermediate-mass} ($\ga$$1.5\msun$) donor stars.

In the past, all binary population synthesis studies that explicitly
considered the evolution of X-ray binaries and the criteria for
dynamically unstable mass transfer involved analytic approximations
\citep[e.g.,][]{rappaport82b,kalogera96a,king99}.  This is a
satisfactory approach as long as the structure of the donor star and
its response to mass loss can be described using relatively simple
prescriptions; however, this is not possible in general.  The clear
and widespread realization that intermediate-mass donor stars can
stably transfer matter to a neutron star (NS) accretor came largely as
a result of recent calculations that utilized full stellar evolution
codes \citep[e.g.,][]{tauris99,podsi00,tauris00,kolb00,podsi02}.  Only
with such codes can the evolution of the donor be followed
realistically during the rapid phase of thermal timescale mass
transfer that characterizes the early evolution of intermediate-mass
X-ray binaries (IMXBs).

\citet[][hereafter, Paper I]{podsi02} is devoted to a systematic
evolutionary study of L/IMXBs, wherein we describe a library of 100
evolutionary sequences computed with a standard Henyey-type stellar
structure code.  This library has now been expanded to 144 sequences,
covering initial orbital periods from 2 hours to 100 days and initial
donor masses from 0.3 to 7$\msun$.  The library is intended to provide
a fairly complete mapping of the initial conditions that are likely to
be encountered in a population synthesis study of L/IMXBs.

Here we extend the work in Paper I by combining our library with a
detailed Monte Carlo binary population synthesis (BPS) code for
L/IMXBs.  The code includes standard assumptions for the population of
massive primordial binaries, reasonable analytic prescriptions to
describe both stable and dynamically unstable mass transfer prior to
the supernova (SN) explosion, and NS kicks.  Similar codes are
described in \citet{zwart96} and \citet{belczynski02}.

We undertake a limited exploration of the set of free parameters that
enter the BPS calculation used to generate the incipient X-ray
binaries.  Probably the most important parameters in the BPS study are
the mean NS kick speed and the envelope binding energy that enters the
prescription for common-envelope evolution.  For reasonable variation
of these two quantities, the formation probability of L/IMXBs ranges
over two orders of magnitude.

For each incipient L/IMXB that emerges from the BPS calculation, we
find an initial model in our library with the closest matching orbital
period and donor mass.  For the ensemble of selected sequences, we
apply a temporal weighting scheme to calculate the distributions of
potentially observable quantities at the current epoch.  This is the
first paper where such distributions have been computed for L/IMXBs,
and it is now possible to directly compare population models and the
statistics of observed systems.


The paper is organized as follows.  In \S~\ref{sec:bps}, we briefly
describe our BPS code, highlighting the important uncertainties and
the associated free parameters.  The population of incipient X-ray
binaries that emerges from the BPS calculation is discussed in
\S~\ref{sec:incip}.  Key results of this study are presented in
\S~\ref{sec:ce}, where we show distributions of various quantities at
the current epoch and make rough comparisons with the observational
data.  Finally, in \S~\ref{sec:con}, the most important results of our
investigation are listed, along with suggestions for how this work may
be extended and improved.



\section{MASSIVE BINARY POPULATION SYNTHESIS}\label{sec:bps}

The formation of a NS in a binary system involves three main
evolutionary steps: (1) the formation of a primordial binary, where
the initially more massive component (the primary) has a mass $\ga 8
\msun$, (2) a phase of mass transfer from the primary to the secondary
(the initially less massive component), and (3) the subsequent SN
explosion of the primary's hydrogen-exhausted core and the formation
of the NS.  Our Monte Carlo BPS code utilizes a set of analytic
prescriptions to describe each of these steps.  A brief overview of
the important elements of our code is given below; an expanded account
is provided in \citet{pfahl02c}.

\subsection{Primordial Binaries}\label{sec:pb}

We construct each primordial binary by selecting the component masses
and orbital parameters from the following distribution functions.

\medskip

\noindent
{\em Primary Mass}.---The initial primary mass, $M_{1i}$, is chosen
from a power-law initial mass function, $p(M_{1i}) \propto
M_{1i}^{-x}$.  We use a fixed value of $x = 2.5$ for massive stars
\citep[e.g.,][]{miller79,scalo86a,kroupa93}.  Primary masses are
restricted to the range $M_{1i} = 8$--$25\msun$, and we assume that
the primary is always the NS progenitor \citep[see,
however,][]{podsi92,pols94,wellstein01}.

\medskip

\noindent
{\em Secondary Mass}.---The initial secondary mass, $M_{2i}$, is
chosen from a distribution in mass ratios, $p(q_i) \propto q_i^y$,
where $q_i \equiv M_{2i}/M_{1i} < 1$.  Strongly motivated by the work
of \citet{garmany80}, we prefer a flat distribution ($y=0$), but we
also consider $y=-1$ and $y=1$.

\medskip

\noindent
{\em Eccentricity}.---Without much loss in generality, we take the
primordial binary orbits to be circular.  This assumption is discussed
in \citet{pfahl02c}.

\medskip

\noindent
{\em Semimajor Axis}.---The initial orbital separation, $a_i$, is drawn
from a distribution that is uniform in $\log a_i$
\citep[e.g.,][]{abt78}.  We determine the minimum value of $a_i$ for
each system by demanding that neither star overflows its Roche lobe on
the main sequence.  The upper limit is somewhat arbitrary, but here is
taken to be $10^3$\,AU.

\subsection{Mass Transfer}\label{sec:mt}

If $a_i\la5$--10\,AU, the primary will grow to fill its Roche lobe at
some point during its evolution.  The subsequent phase of mass
transfer is of crucial importance in determining what types of NS
binaries are ultimately produced.  It is common practice to
distinguish among three main evolutionary phases of the primary at the
onset of mass transfer \citep{kippenhahn67,podsi92}.  Case A evolution
corresponds to core hydrogen burning, case B refers to the shell
hydrogen-burning phase, but prior to central helium ignition, and case
C evolution begins after core helium burning.  It is quite improbable
for mass transfer to begin during core helium burning, and we thus
neglect this possibility \citep[see][]{pfahl02c}.  We refer to as case
D the large fraction of wide binaries that remain detached prior to
the SN explosion of the primary.  Using the distributions and
standard-model parameters given above, as well as the treatment of
stellar winds discussed below, we find that cases A, B, C, and D
comprise roughly 5\%, 30\%, 15\%, and 50\%, respectively, of the
primordial binary population.  In order to determine which case each
binary falls into, we use the single-star evolution fitting formulae
of \citet{hurley00}.

Mass transfer from the primary to the secondary may be {\em stable} or
{\em dynamically unstable}, depending mainly on the binary mass ratio
and evolutionary state of the primary when it fills its Roche lobe.
In our population study of L/IMXBs, we consider only cases B and C
mass transfer.  Case A mass transfer accounts for only a small
fraction of the binaries and, furthermore, most likely leads to the
merger of the two stars following a contact phase
\citep[e.g.,][]{wellstein01}. Most case D systems are disrupted due to
the SN explosion of the primary if NS kicks are significant.  We do
not consider case D systems that survive the SN; for a discussion of
the products that may emerge from this evolutionary channel, see
\citet{kalogera98b} and \citet{willems02}.

Cases B and C are divided in {\em early} (B$_e$ or C$_e$) and {\em
late} (B$_l$ or C$_l$) phases, if the primary has an envelope that is
mostly radiative or deeply convective, respectively.  We assume that
mass transfer is stable, though non-conservative, for cases B$_e$ and
C$_e$ if the mass ratio, after any wind mass loss has occurred, is
$>$$q_c$, where $q_c$ is some critical mass ratio.  We adopt a fixed
value of $q_c = 0.5$ in our study \citep[e.g.,][]{wellstein01}.  If
the primary has a deep convective envelope when it fills its Roche
lobe, mass transfer is dynamically unstable and a common-envelope (CE)
phase ensues, which results in either a very compact binary or a
merger.

A single star of mass $\ga$$15\msun$ may lose $\ga$30\% of its mass in
a stellar wind on the asymptotic giant branch (AGB).  For stars of
mass $\la$$25\msun$, only $\la$5\% of the mass is lost on the main
sequence.  We suppose that the wind from the primary in a binary
system takes with it the specific orbital angular momentum of the
star.  In response to the AGB winds, the Roche lobe of the primary
expands and may overtake the expansion of the star, making Roche-lobe
overflow and case C$_l$ mass transfer impossible.  We have included
the effects of stellar winds only for initial primary masses
$>$$13\msun$, on both the main sequence and the AGB; our procedure is
similar to the one adopted by \citet{podsi03}.  For this range of
masses, core helium burning begins while the star is in the
Hertzsprung gap, and there is no decrease in the stellar radius.
Primaries of mass $M_{1i}\la 13\msun$ experience moderate wind mass
loss during core helium burning following evolution through the first
giant branch, but the stellar radius decreases after helium ignition,
precluding Roche-lobe overflow during this phase.  We note that the
mass that separates the two behaviors just mentioned is actually quite
uncertain \citep[e.g.,][]{langer95}, and may be as large as
$\sim$$20\msun$.

If a merger is avoided, it is reasonable to suppose that the primary
loses its entire envelope, leaving only its hydrogen-exhausted core,
irrespective of whether mass transfer is stable or dynamically
unstable.  Following case B mass transfer, the mass of the helium core
is given approximately by \citep{hurley00}
\be\label{eq:core}
M_{ci} \simeq 0.1 M_{1i}^{1.35}~.
\ee 
We neglect the relatively small amount of wind mass loss on the main
sequence and use the initial mass of the primary to compute the
initial core mass $M_{ci}$ before it too loses mass in a wind.  The
mass of the exposed core may be larger by $\sim$0.5--$1\msun$ after
case C mass transfer, as a result of shell nuclear burning.  A helium
core will ultimately yield a NS remnant if its mass is $\ga$$2\msun$
\citep{habets86b}.  Equation~(\ref{eq:core}) gives a primary mass
threshold for NS formation of $\sim$$9\msun$

For our chosen maximum primary mass of $25\msun$, the corresponding
core mass is $\sim$$8\msun$.  A nascent helium star of mass
3--$8\msun$ that is exposed following case B mass transfer may lose
10--30\% of its mass in a wind before the SN
\citep[e.g.,][]{brown01,pols02}.  The final core mass is related to
the initial helium star mass by the approximate formula \citep[see
Fig. 1 of][]{pols02}
\be
M_{cf} \sim 1.4 M_{ci}^{2/3}~,
\ee
for $M_{ci} \ga 3\msun$.  Following case C mass transfer, the core of
the primary has already undergone helium burning, and there is
insufficient time for winds to significantly reduce its mass.  

If $M_{ci} \la 3\msun$ following case B mass transfer, we may safely
neglect winds, but such helium stars may expand to giant dimensions
following core helium burning, often initiating a phase of so-called
case BB mass transfer to the secondary
\citep{degreve77,delgado81,habets86a}.  We do not attempt to model
this evolution in detail, but simply assume that $0.5\msun$ is
transferred conservatively from the primary's core to the secondary.
In case BB systems where $M_{2i} < M_{ci}$, such as when the secondary
is of low- or intermediate-mass, the mass transfer may proceed on the
thermal timescale of the core, and the evolution may be quite
complicated \citep[e.g.,][]{dewi02,ivanova03}.  However, any
reasonable treatment of case BB mass transfer is not likely to change
our results for L/IMXBs substantially,

Stable mass transfer from the primary to the secondary is treated
analytically as follows.  We assume that the secondary accretes a
fraction $\beta$ of the material lost from the primary during
Roche-lobe overflow.  The complementary mass fraction, $1-\beta$,
escapes the system with specific angular momentum $\alpha$, in units
of the orbital angular momentum per unit reduced mass.  We use
constant values of $\alpha = 1.5$, characteristic of mass loss through
the L2 point, and $\beta = 0.75$.  The final orbital separation is
then given by the generic equation \citep{podsi92}
\be\label{eq:albe}
\left( \frac{a'}{a} \right)_{\rm RLO} = 
\frac{M_b'}{M_b} \left( \frac{M_1'}{M_1} \right)^{C_1}
\left( \frac{M_2'}{M_2} \right)^{C_2} ~,
\ee
where
\bea
C_1 & \equiv & 2\alpha(1-\beta)-2 \nonumber \\ 
C_2 & \equiv & -2\alpha(1-\beta)/\beta-2 ~.
\eea
Here the subscript `RLO' denotes stable Roche-lobe overflow, and
unprimed and primed quantities denote parameters at the onset and
termination of mass transfer, respectively.

It is easily verified that, in our simulations, the {\em minimum}
secondary mass resulting from stable mass transfer is roughly
$q_c\,8\msun + \beta\,6\msun$, where $8\msun$ is the minimum primary
mass and $6\msun$ is the corresponding envelope mass shed during mass
transfer.  For our chosen values of $q_c = 0.5$ and $\beta = 0.75$,
this minimum mass is $8.5\msun$, considerably larger than the maximum
initial donor mass of $\sim$$4\msun$ for which an IMXB undergoes
stable mass transfer (see Paper I).  Thus, in our work, all incipient
L/IMXBs are the products of dynamically unstable mass transfer.

We use the conventional energy relation to describe the dynamical
spiral-in during a CE phase \citep[e.g.,][]{webbink84}.  The ratio of
the final to the initial orbital separation is given by the generic
equation
\be\label{eq:ce}
\left( \frac{a'}{a} \right)_{\rm CE} = \frac{M_c M_2}{M_1} 
\left( M_2 + \frac{2 M_e}{\lambda_{\rm CE} \, \eta_{\rm CE} 
\, r_{\rm L1}} \right)^{-1} ~,
\ee
where $r_{\rm L1}$ is the Roche-lobe radius of the primary in units of
the orbital separation, $M_e$ and $M_c$ are the masses of the
primary's envelope and core, respectively, $\lambda_{\rm CE}$
parameterizes the structure of the envelope, and $\eta_{\rm CE}$ is
the fraction of orbital binding energy that goes into dissipating the
envelope.  Although $\lambda_{\rm CE}$ and $\eta_{\rm CE}$ appear only
in the product $\lambda_{\rm CE}\,\eta_{\rm CE}$, we suppose that
$\eta_{\rm CE} = 1$ and fix $\lambda_{\rm CE}$ in the range 0.1--0.5
for each simulation.  In reality, $\lambda_{\rm CE}$ changes as a star
evolves, typically decreasing to $\la$0.1--0.2 as the massive star
passes through the Hertzsprung gap, and increasing to $\sim$0.1--0.4
as the star ascends the AGB \citep{dewi00,dewi01}.  Generally,
$\lambda_{\rm CE}$ decreases for all stellar radii as the mass of the
star is increased.  Equation~(\ref{eq:ce}) gives typical shrinkage
factors of $(a'/a)_{\rm CE} \sim 0.01$ for systems with low- and
intermediate-mass secondaries.

A sufficient condition for the merger of the primary and secondary is
that the main-sequence secondary overfills its Roche lobe for the
calculated post-CE orbital separation.  Therefore, the minimum
separation for surviving systems must be larger than several solar
radii, corresponding to initial orbital separations greater than
several {\em hundred} solar radii.  It turns out that the majority of
the dynamically unstable case B$_e$ and C$_e$ systems merge following
the CE, and that in most systems that survive the CE, the primary is a
convective red supergiant (case B$_l$ or C$_l$) at the onset of mass
transfer.

\subsection{Supernova Explosion}\label{sec:sn}

After the exposed core of the primary consumes its remaining nuclear
fuel, it explodes as a Type Ib or Ic SN and leaves a NS remnant.  We
take the initial NS mass to be $1.4\msun$.  Impulsive mass loss and
the NS kick strongly perturb the binary and may cause its disruption.
Mass loss is especially significant, since the mass ejected may be
comparable to or greater than the secondary mass. Some important
insights can be obtained rather simply by neglecting NS kicks.

It is straightforward to show \citep[e.g.,][]{blaauw61,boersma61} that
for a circular pre-SN orbit and {\em vanishing kicks} the
eccentricity, $e_{\rm SN}$, after the SN is simply
\be\label{eq:eccsn}
e_{\rm SN} = \frac{M_c - M_{\rm NS}}{M_2 + M_{\rm NS}}~,
\ee
where $M_{\rm NS}$ is the mass of the NS, and $M_c$ is the
pre-explosion core mass.  The system is unbound when $M_c > M_2 + 2
M_{\rm NS}$.  Letting $M_2 = 1\msun$, we see that for $M_c > 3.8\msun$
($M_1 \ga 15\msun$) disruption of the binary is guaranteed.  A kick of
appropriate magnitude and direction is then {\em required} in order to
keep the system bound.  For intermediate-mass secondaries, a wider
range of pre-SN core masses is permitted when kicks are neglected.
Equation~(\ref{eq:eccsn}) gives values of $e_{\rm SN} \ga 0.5$ for
some typical core and secondary masses.  Following the CE phase, the
core of the primary will have orbital speeds about the binary
center-of-mass (CM) of $v_c \ga 200\kms$.  After the SN, mass loss
alone gives the binary CM a speed of $v_{\rm CM} = e_{\rm SN} v_c$, so
that eccentricities of $e_{\rm SN} \ga 0.5$ correspond to large
post-SN systemic speeds of order $100\kms$.

When NS kicks are considered in addition to SN mass, a larger fraction
of systems are disrupted, and those binaries that do remain bound will
have larger CM speeds.  We utilize a Maxwellian distribution in kick
speeds,
\be\label{eq:kick}
p(v_k) = \sqrt{\frac{2}{\pi}} \frac{v_k^2}{\sigma_k^3} 
\exp (-v_k^2/2\sigma_k^2)~, 
\ee
where the directions of the kicks are distributed isotropically.
Dispersions of $\sigma_k \sim 100$--$200\kms$ are reasonably
consistent with the data on pulsar proper motions
\citep[e.g.,][]{hansen97,arzoumanian02}.  However, neither the
functional form of the kick distribution nor the mean are very well
constrained.  In our study, we consider $\sigma_k = 50$, 100, and
$200\kms$.  The post-SN orbital parameters are calculated using the
formalism described in Appendix~B of \citet{pfahl02c}.

Significant SN mass loss and large NS kicks yield bound post-SN
binaries with high eccentricities.  Given the post-SN eccentricity, we
check if the radius of the secondary is $>$10\% larger than its tidal
radius at periastron.  If this occurs, we assume that the NS
immediately spirals into the envelope of the unevolved secondary and
do not consider the system further.  Our 10\% overflow restriction
allows for the possibility that tidal circularization and perhaps some
mass loss will prevent the objects from merging.  The details of
eccentric binary evolution with mass transfer is well beyond the scope
of this investigation.  However, we note in passing that Cir X-1 is
most likely a young, possibly intermediate-mass, X-ray binary
undergoing episodic mass transfer as a result of having a highly
eccentric ($e_{\rm SN} \ga 0.8$) orbit \citep{shirey98}.

If the coalescence of the NS and secondary is avoided, we neglect mass
loss from the system and assume, rather simplistically, that the
binary circularizes while conserving orbital angular momentum.  The
final orbital separation is then $a_{\rm SN}(1-e_{\rm SN}^2)$, where
$a_{\rm SN}$ is the semimajor axis after the SN.  Moreover, we assume
that the secondary rotates synchronously with the circularized orbit.


\section{INCIPIENT X-RAY BINARIES}\label{sec:incip}

The output of the population synthesis calculation is a set of
circular binaries, each identified by their orbital period and the
mass of the secondary.  In order to select initial models from our
library of L/IMXB evolutionary sequences, we require the orbital
period at which the secondary first fills its Roche lobe. Hereafter,
we denote the donor and accretor (NS) masses in units of $M_\odot$ by
$M_d$ and $M_a$, respectively.

Orbital angular momentum losses via gravitational radiation (GR) and
magnetic braking (MB) may cause the binary separation to decrease
substantially prior to mass transfer.  The timescale for orbital
shrinkage due to GR is
\be\label{eq:gr}
\tau_{\rm GR} 
\simeq 110\,{\rm Gyr}\,\,q^{-1}(1+q)^{1/3}
\left(\frac{P_{\rm orb}}{1\,{\rm day}}\right)^{8/3}~,
\ee
where $q = M_d/M_a$, and $M_a = 1.4$.  We utilize the same standard MB
formula as in Paper I \citep{verbunt81,rappaport83b}, with a
characteristic timescale given by
\be\label{eq:mb}
\tau_{\rm MB} \simeq
3.5\,{\rm Gyr}\,\, 
\eta_{\rm MB}^{-1} (1+q)^{-1/3} R_d^{-4} 
\left(\frac{P_{\rm orb}}{1\,{\rm day}}\right)^{10/3}~,
\ee
where $R_d$ is the radius of the secondary in solar units, and we
introduce $\eta_{\rm MB}$ to parameterize the strength of MB prior to
mass transfer.  Only when the secondary is tidally locked to the orbit
does MB extract orbital angular momentum.  In an ad hoc way, we let
$\eta_{\rm MB} = 0$ or 1 to model the cases where the tidal coupling
prior to Roche-lobe overflow is very weak or very strong,
respectively.  We apply MB only if $0.4 < M_d < 1.5$; less massive
stars are fully convective and may not undergo MB
\citep{rappaport83b}, while more massive stars have radiative
envelopes and may not undergo MB.  When the main sequence lifetime of
a low-mass star exceeds the $\simeq$13\,Gyr age of the Galaxy, MB or
GR are {\em required} to drive the system into contact.
Intermediate-mass secondaries have nuclear lifetimes of
$\sim$0.1--3\,Gyr, and so do not suffer from this obstacle.


The single-star evolution code of \citet{hurley00} is used to follow
the radial evolution of the secondary as MB and GR shrink the orbit.
The metallicity is set to the solar value of $Z = 0.02$.  We neglect
the evolutionary time prior to the SN of the primary and assume that
the secondary is on the ZAMS at the time the NS is formed.  The age of
the secondary when it fills its Roche lobe is here referred to as the
``lag time,'' denoted by $t_{\rm lag}$.  The lag time thus
approximates the time between the formation of the primordial binary
and the onset of the X-ray binary phase.  We take the age of the
Galaxy to be 13\,Gyr, and only accept as incipient L/IMXBs those for
which $t_{\rm lag} < 13\gyr$.

Figure~\ref{fig:incip} shows an illustrative set of distributions of
binary parameters and systemic speeds for the incipient L/IMXBs.
Overlayed on the scatter plots are the initial models ({\em open
circles}) from our library of L/IMXBs evolutionary sequences.  We
applied fixed values of $y=0.0$, $\sigma_k = 200\kms$, $\eta_{\rm MB}
= 1$, and considered CE structure parameters of $\lambda_{\rm CE} =
\{0.1, 0.3, 0.5\}$.  For each parameter set shown, the distribution of
initial secondary masses increases with mass.  This distribution is
modified if we adopt a different distribution of mass ratios for the
components of the primordial binaries.  Nonetheless, the general
statistical importance of initially intermediate-mass secondaries is
clear.  The high concentration of systems with $P_{\rm orb} \la
0.5\day$ and $M_d \la 1.5$ when $\lambda_{\rm CE} = 0.3$ and 0.5 is
due to MB and GR.  The absence of LMXBs with evolved secondaries of
mass $M_d \la 1$ for $P_{\rm orb} \ga 0.5\day$ when $\lambda_{\rm CE}
= 0.3$ and 0.5 results from the demand that $t_{\rm lag} < 13\gyr$.
More low-mass donors survive this cut if the metallicity is reduced,
since the main-sequence lifetime decreases with decreasing
metallicity.  Note that for $\lambda_{\rm CE} = 0.1$ no systems with
low-mass donors survive the CE, and that the maximum period of the
incipient IMXBs is only $\sim$3\,day.  When $\lambda_{\rm CE} = 0.1$,
the small fraction of incipient IMXBs with $2.5 < M_d < 4$ is not well
sampled by our grid of evolutionary sequences, but we expect that a
denser sampling will not greatly change our results.

It is somewhat interesting that systems with fully convective
secondaries of mass $M_d \la 0.4$ can be driven into contact by GR
within the lifetime of the Galaxy (eq.~[\ref{eq:gr}]).  A handful of
such binaries are present in Fig.~\ref{fig:incip} for $\lambda_{\rm
CE} = 0.3$ and 0.5.  Mass transfer will be driven by GR at rates of
$\la$$10^{-10}\,M_\odot\yr^{-1}$, from initial periods of
$\sim$2--3\,hr to a minimum period of $\sim$1.5\,hr, at which point
$M_d \sim 0.05$.  The secondary and orbit then expand in response to
further mass transfer.  For a physical discussion of LMXB evolution
with very low-mass donors and the period minimum, see
\citet{nelson86}.  The companion mass ($M_d \ga 0.04$) and orbital
period ($P_{\rm orb} \simeq 2$\,hr) of the first-discovered
millisecond X-ray pulsar SAX~J1808.4--3658
\citep{wijnands98,chakrabarty98} are consistent with it having formed
and evolved in the way just described.  However, given the uncertainty
in the binary inclination, a model wherein the initial donor mass
takes a more typical value of $\sim$$1\msun$ is also consistent with
the present system parameters (L. Nelson \& S. Rappaport 2003, in
preparation).

\begin{figure*}
\centerline{\epsfig{file=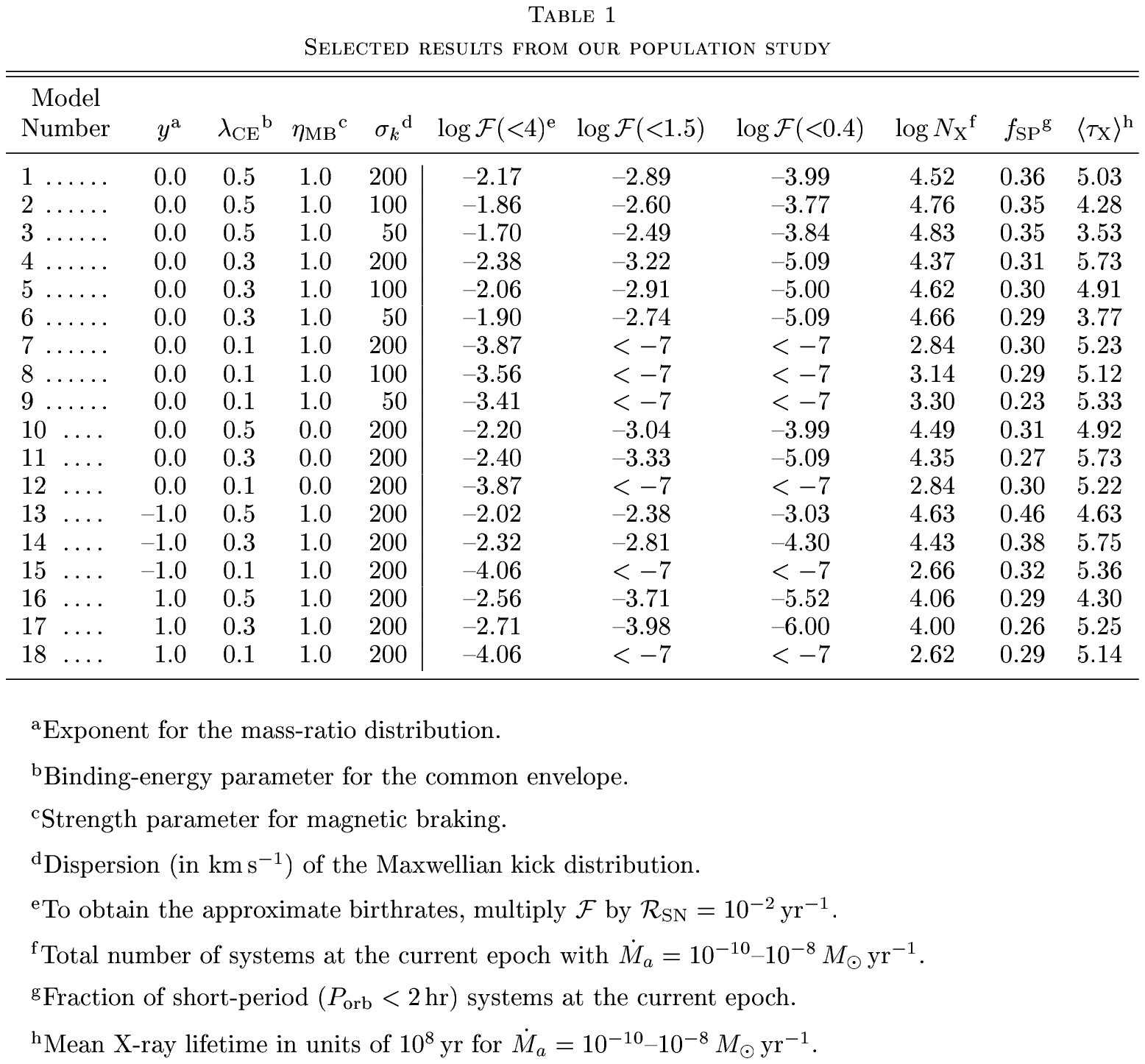,width=0.7\linewidth}}
\end{figure*}

We define the {\em formation efficiency} as $\mathcal{F}(<\!\!\!M_d) =
N_{\rm inc}(<\!\!\!M_d)/N_{\rm PB}$, where $N_{\rm inc}(<\!\!\!M_d)$
is the synthesized number of incipient L/IMXBs with donor masses less
than $M_d$, and $N_{\rm PB}$ is the number of primordial binaries used
in the simulation.  The majority of systems with $M_d > 4$ undergo
dynamical mass transfer shortly after Roche-lobe overflow (see Paper
I).  Therefore, we take $\mathcal{F}(<\!\!4)$ to be the {\em total}
formation efficiency of LMXBs and IMXBs.  We somewhat arbitrarily
define incipient LMXBs as systems with initial donor masses of $M_d <
1.5$.  The present rate of core-collapse SNe in the Milky Way is
$\mathcal{R}_{\rm SN} \sim 10^{-2}\yr^{-1}$
\citep[e.g.,][]{cappellaro99}.  Here we adopt the simplifying
assumption that the average star formation rate has been constant for
13\,Gyr and take $\mathcal{R}_{\rm SN}$ to be the approximate
formation rate of massive primordial binaries.  It follows that the
current Galactic birthrate of L/IMXBs is $\mathcal{B}(<\!\!\!M_d) =
\mathcal{F}(<\!\!\!M_d) \mathcal{R}_{\rm SN}$.

Table~1 lists formation efficiencies for a modest number of different
parameter sets in the BPS study, where we vary only $y$, $\lambda_{\rm
CE}$, $\sigma_k$, and $\eta_{\rm MB}$.  We used $10^7$ primordial
binaries in each simulation.  For the parameter sets explored, $\log
\mathcal{F}(<\!\!4)$ varies from roughly $-4.1$ to $-1.7$, with
corresponding birth rates of $\mathcal{B}(<\!\!4) \sim
10^{-6}$--$10^{-4}\yr^{-1}$.  Incipient LMXBs are formed with
systematically lower efficiencies of $\log \mathcal{F}(<\!\!1.5) \la
-2.4$, and birthrates of $\mathcal{B}(<\!\!1.5) \la 4\times
10^{-5}\yr^{-1}$.  The range of LMXB birthrates is reasonably
consistent with the results of, e.g., \citet{zwart96} and
\citet{kalogera98a}, each of whom used somewhat different assumptions
than the ones adopted here.  The formation efficiencies for $M_d <
0.4$ are also listed in Table 1, with typical values of $\log
\mathcal{F}(<\!\!0.4) \la -4$.  If SAX~J1808.4--3658 evolved in the
way described above, then $\lambda_{\rm CE} \sim 0.5$ if favored,
since this value yields the largest birthrates of
$\mathcal{B}(<\!\!1.5) \sim 10^{-6}\yr^{-1}$.

\begin{figure*}
\begin{minipage}[t]{0.41\linewidth}
\centerline{\epsfig{file=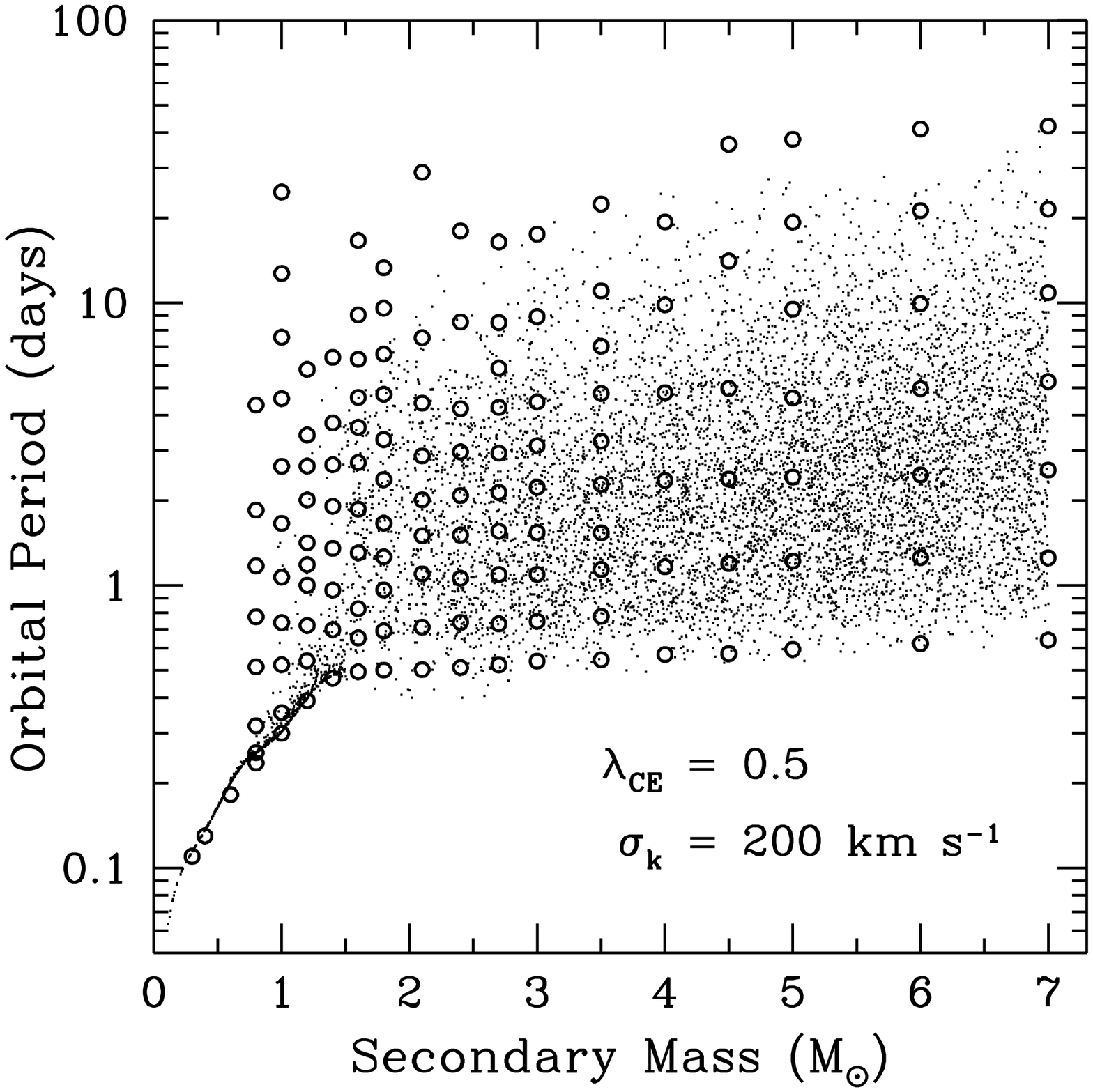,width=0.9\linewidth}}
\end{minipage}
\hspace{0.4cm}
\begin{minipage}[t]{0.41\linewidth}
\centerline{\epsfig{file=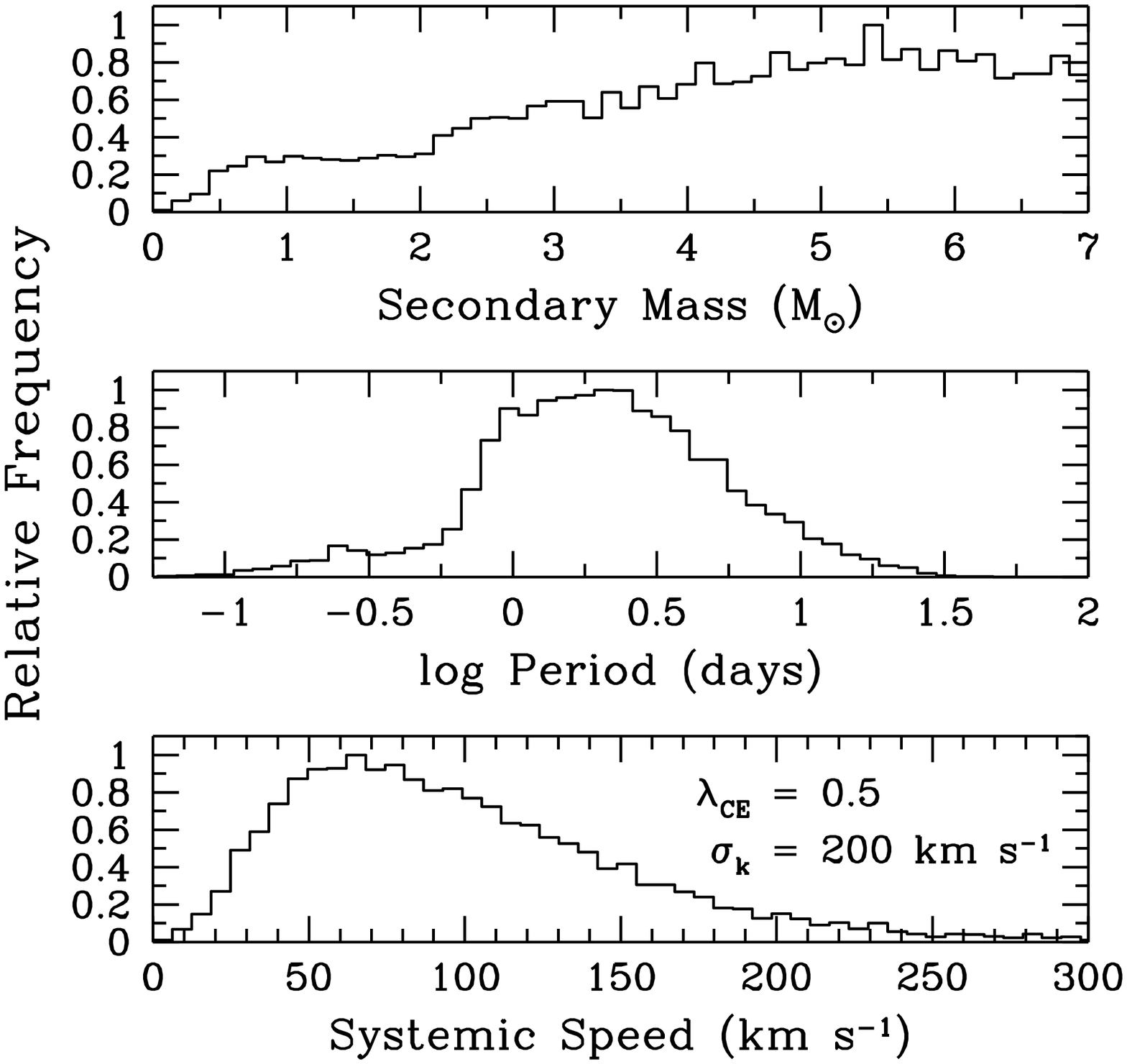,width=0.9\linewidth}}
\end{minipage} \vspace{5mm}
\\
\begin{minipage}[t]{0.41\linewidth}
\centerline{\epsfig{file=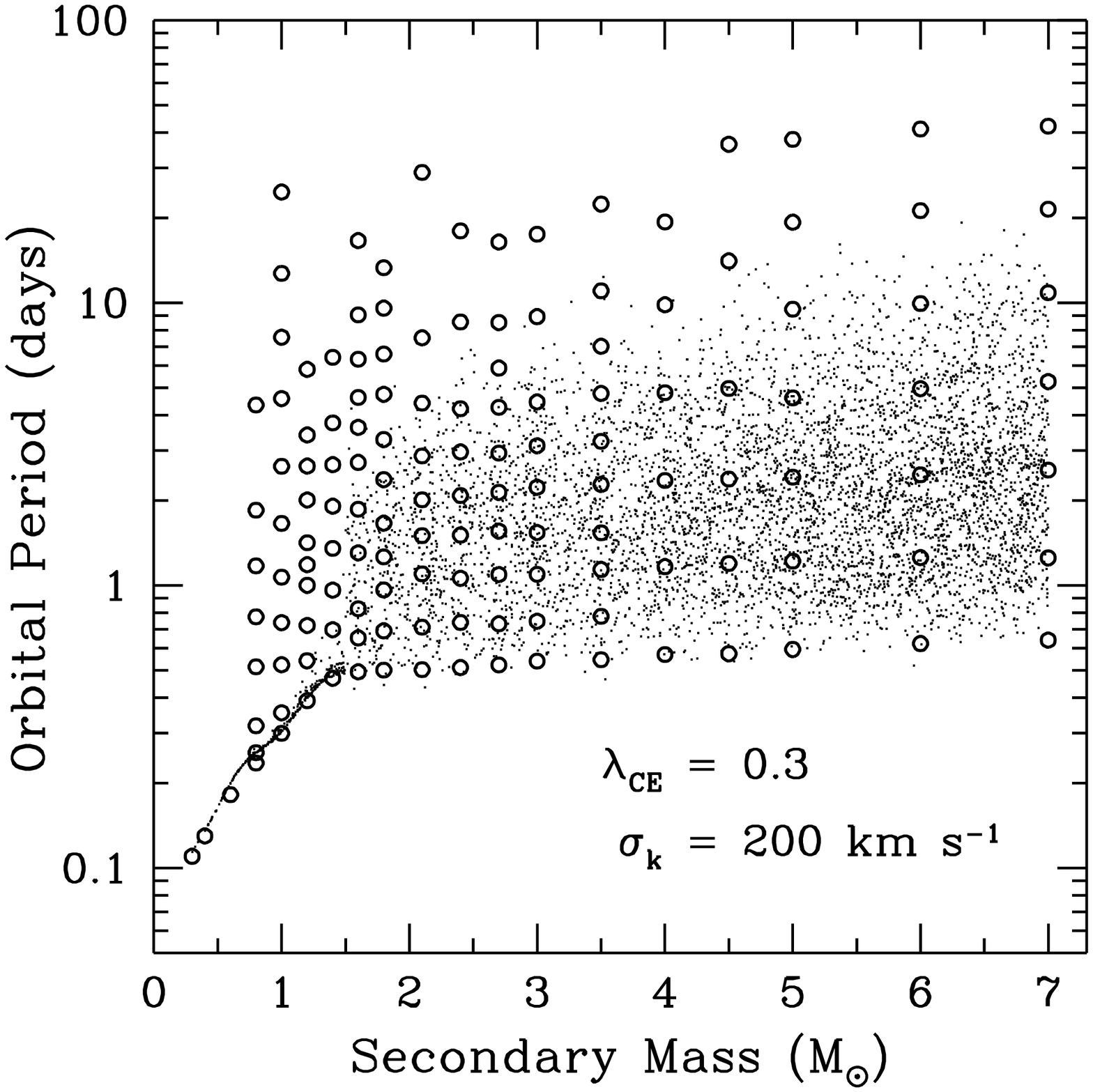,width=0.9\linewidth}}
\end{minipage}
\hspace{0.4cm}
\begin{minipage}[t]{0.41\linewidth}
\centerline{\epsfig{file=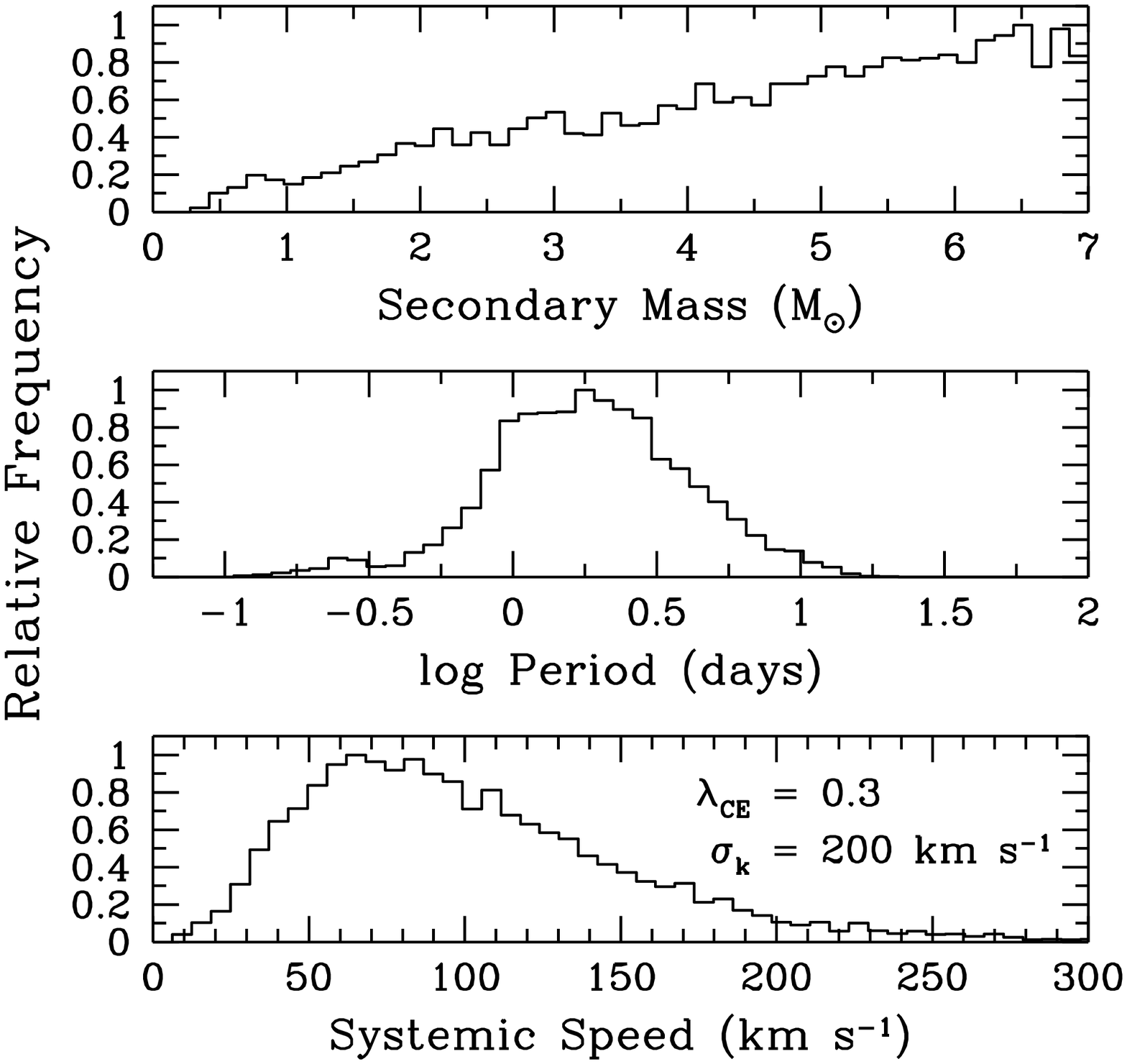,width=0.9\linewidth}}
\end{minipage} \vspace{5mm}
\\ 
\begin{minipage}[t]{0.41\linewidth}
\centerline{\epsfig{file=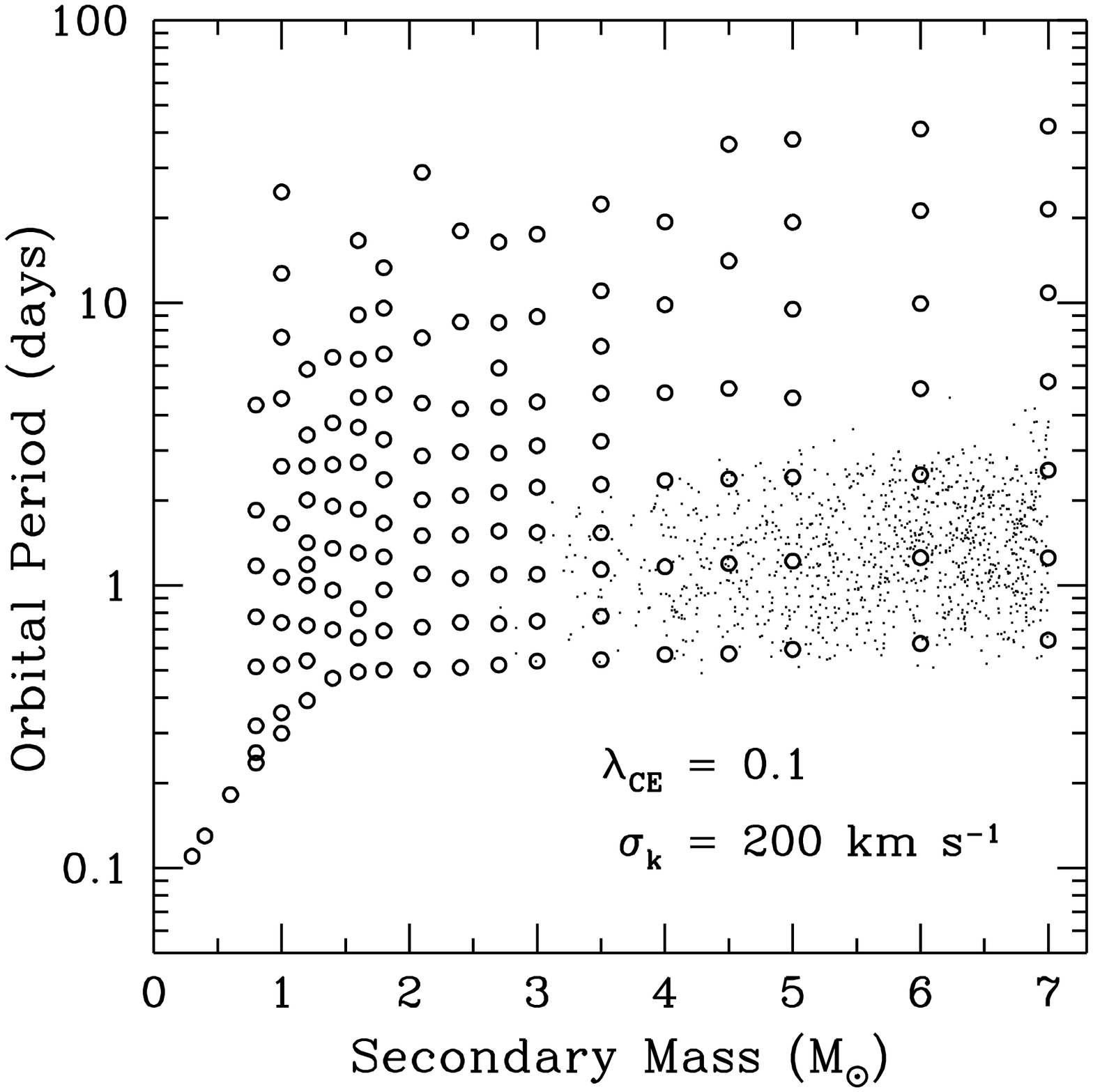,width=0.9\linewidth}}
\end{minipage}
\hspace{0.4cm}
\begin{minipage}[t]{0.41\linewidth}
\centerline{\epsfig{file=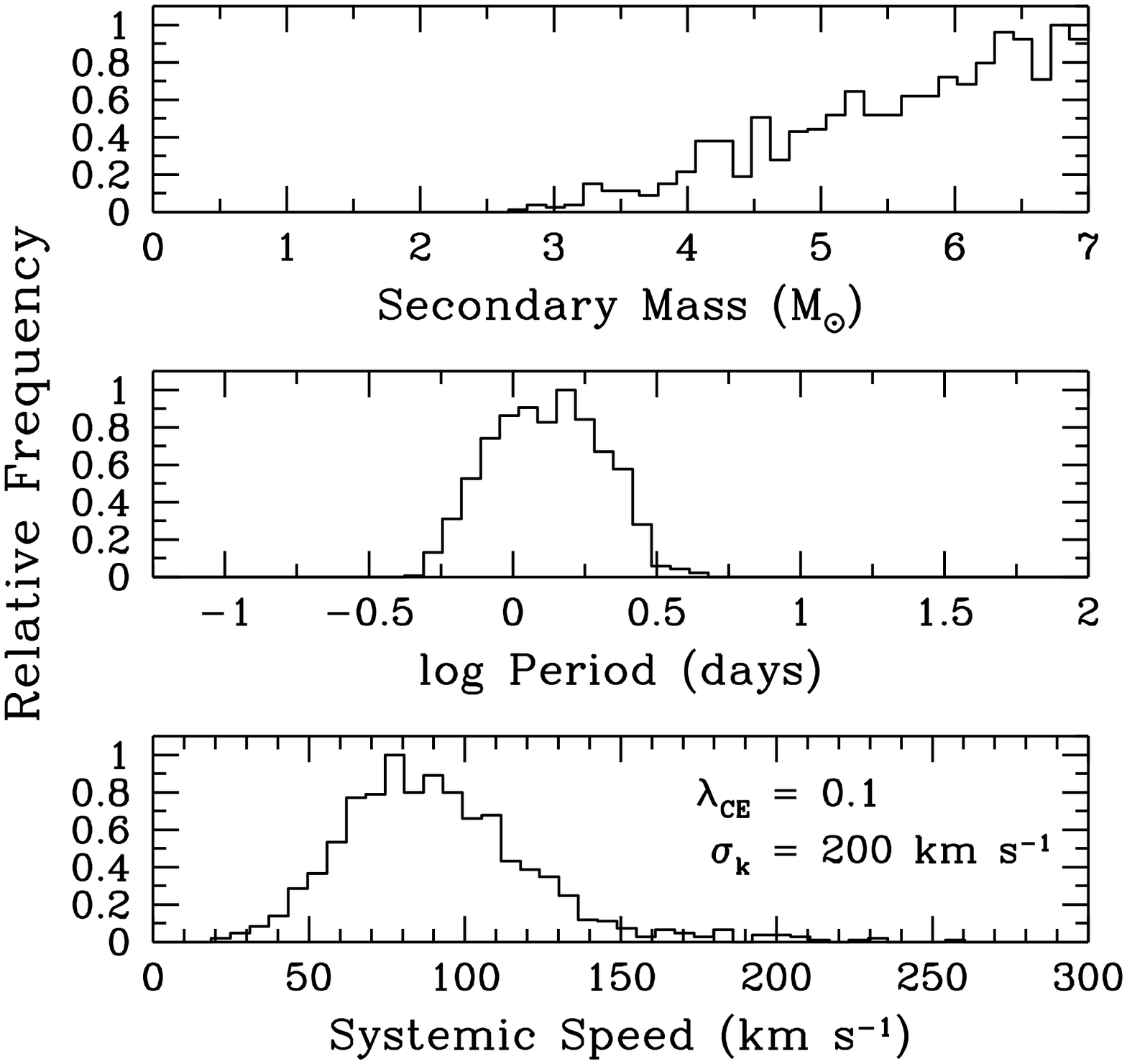,width=0.9\linewidth}}
\end{minipage}
\caption{Scatter plots and histograms for incipient L/IMXBs.  For
these simulations we used 500,000 primordial binaries and parameter
values of $y= 0.0$, $\sigma_k = 200\kms$, $\eta_{\rm MB} = 1$, and
$\lambda_{\rm CE} = \{0.1, 0.3, 0.5\}$.  The clustering of systems
with $P_{\rm orb} \la 0.5\day$ and $M_d \la 1.5$ for $\lambda_{\rm CE}
= 0.3$ and 0.5 is due to the effects of magnetic braking and
gravitational radiation. The histograms also show the distributions
of systemic speeds that result from the SNe.  For reference, we have
overlayed on the scatter plots the initial models in our library of
evolutionary sequences ({\em open circles}).
\label{fig:incip}}
\end{figure*}


\section{POPULATION AT THE CURRENT EPOCH}\label{sec:ce}

In order to compare our theoretical population synthesis results and
the statistics of observed systems, it is necessary to follow the
evolution of each L/IMXB that we generate.  For each synthesized
system, a close match is found in our library of evolutionary
sequences.  Distributions at the current epoch of observable
quantities are computed by appropriately weighting each evolutionary
sequence.  We now elaborate on these points and discuss several key
results, as well as make very rudimentary comparisons between our
results and the observations.

\subsection{Selection from the Library}

Each incipient L/IMXB is characterized by the donor mass and the
circularized orbital period at which the star first fills its Roche
lobe.  For any given initial $P_{\rm orb}$ and $M_d$, we select a
model from our L/IMXB evolutionary library by first identifying the
subset of sequences in the library with the closest initial donor
mass, and then finding the one sequence in this subset with the
closest initial orbital period.  Incipient LMXBs with $M_d < 0.3$ are
evolved by selecting the one sequence with an initial donor mass of
$0.3\msun$ and considering only the part of the sequence for which the
donor mass is $<$$M_d$.  This is reasonable, since such stars will
initially follow the approximate mass-radius relation, $R_d \propto
M_d^{0.8}$, for low-mass stars in thermal equilibrium.  If the library
sequences are labeled with some integer index $i$, we may define the
formation efficiency $\mathcal{F}_i$ of sequence $i$ as the number of
times that sequence is selected, divided by $N_{\rm PB}$.

\subsection{Weighting Procedure}

For each selected library model, the entire evolutionary sequence
contributes to the distributions at the current epoch of $P_{\rm
orb}$, $M_d$, and $\dot{M}_a$, the accretion rate onto the NS.  This
is done as follows. Let $Q$ be the quantity of interest.  The
evolutionary data file for each sequence gives $Q$ as a function of
the time $t_{\rm MT}$ since the onset of mass transfer to the NS.  For
some small interval of time $\delta t_{\rm MT}$, $Q$ varies over a
small range $(Q', Q' + \delta Q)$.  The probability that there is an
identical system that is {\em presently} in this particular state
somewhere in the Galaxy is $\mathcal{F}_i \mathcal{R}_{\rm SN}(t_{\rm
PB}) \delta t_{\rm MT}$, where $t_{\rm PB} = - (t_{\rm MT} + t_{\rm
lag})$ is the formation time of the primordial binary, and {\em now}
is taken to be at time $t = 0$.  The total number of L/IMXBs present
in the Galaxy with $Q$ in the bin $(Q_0, Q_0 + \Delta Q)$ is $\sum
\mathcal{F}_i \mathcal{R}_{\rm SN}(t_{\rm PB}) \delta t_{\rm MT}$.
Here the symbolic sum is over all sequences $i$ and all evolutionary
times for which $Q$ lies in the chosen bin.  While the above procedure
is valid for a variable star formation rate, we limit our study to a
constant massive binary formation rate of $\mathcal{R}_{\rm SN} =
10^{-2}\,{\rm yr}^{-1}$ (see \S~\ref{sec:incip}).

\begin{figure*}
\begin{minipage}[t]{0.5\linewidth}
\centerline{\epsfig{file=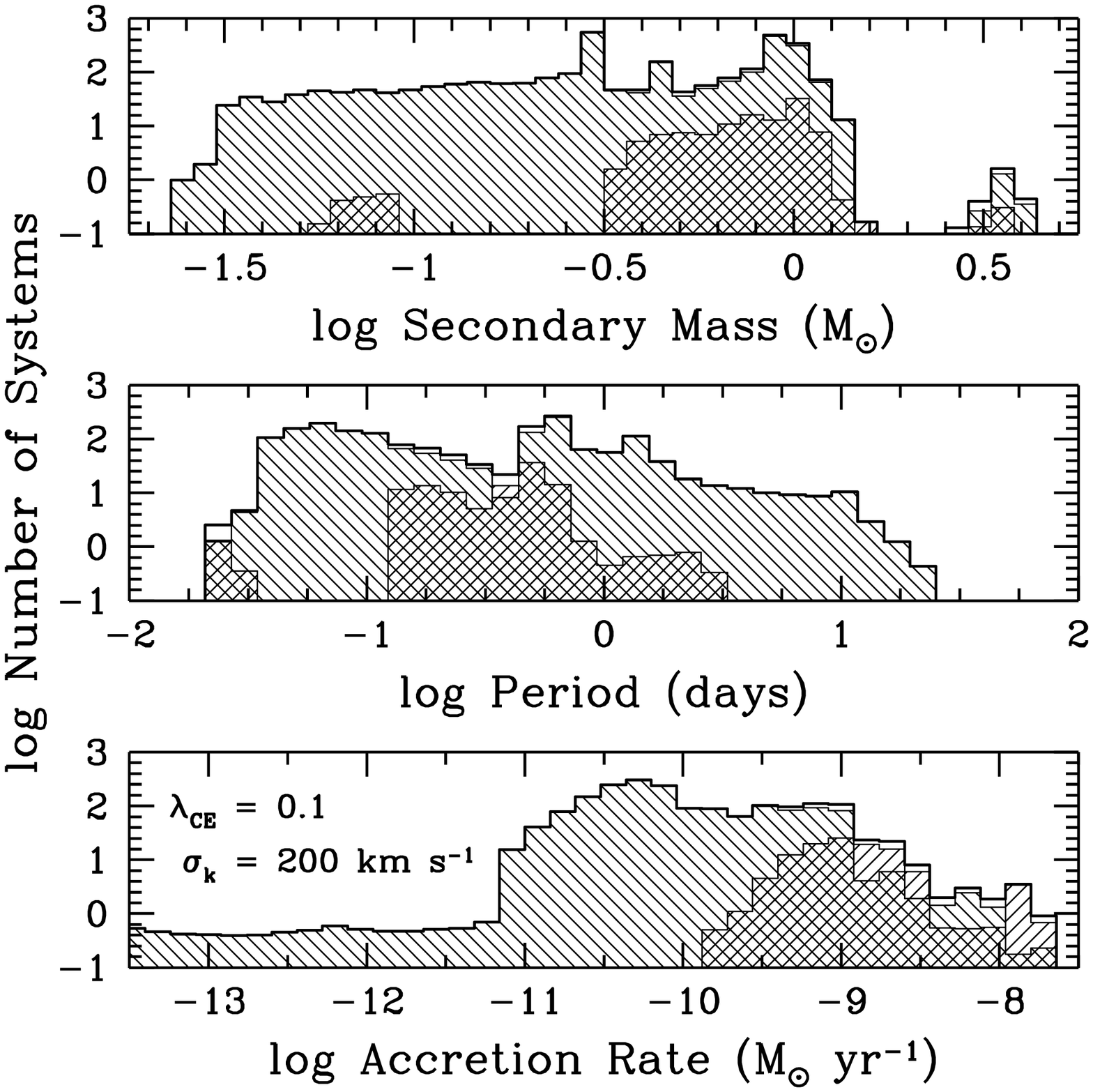,width=0.9\linewidth}}
\end{minipage}
\hfill
\begin{minipage}[t]{0.5\linewidth}
\centerline{\epsfig{file=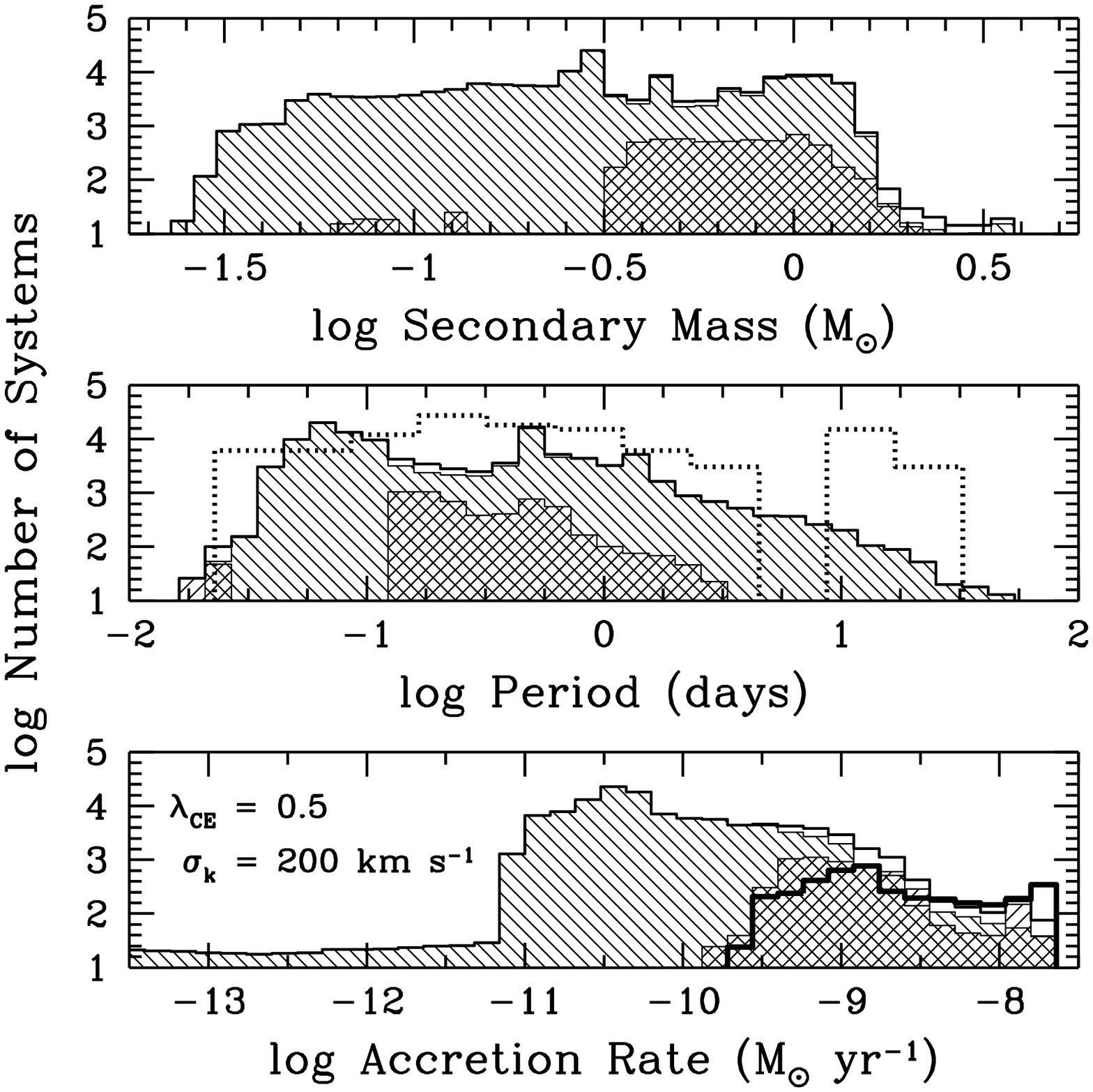,width=0.9\linewidth}}
\end{minipage}
\caption{Distributions at the current epoch of the orbital period,
donor mass, and mass accretion rate onto the NS, for parameters $y =
0.0$, $\sigma_k = 200\kms$, $\eta_{\rm MB} = 1$, and $\lambda_{\rm CE}
= 0.1$ ({\em left}) and 0.5 ({\em right}).  We have assumed a constant
massive binary formation rate of $10^{-2}\yr^{-1}$. Note the
difference in scale between the left and right figures.  The hatched
regions indicate persistent ($+45\degr$) and transient ($-45\degr$)
X-ray sources, and the enclosing solid histogram gives the sum of
these two populations.  Overlayed (dotted histogram) on the
theoretical period distribution in the figure on the right is the
rescaled distribution of 37 measured periods \citep{liu01} among
$\sim$140 observed LMXBs in the Galactic plane.  The thick, solid
distribution overlayed on the bottom panel of the right figure
illustrates how the inclusion of X-ray irradiation effects might
change our theoretical $\dot{M}_a$ distribution.
\label{fig:cedis}}
\end{figure*}

\subsection{Current-Epoch Distributions}

Figure~\ref{fig:cedis} shows theoretical distributions (solid
histograms) at the current epoch of $M_d$, $P_{\rm orb}$, and
$\dot{M}_a$, where we have used illustrative values of $y = 0.0$,
$\sigma_k = 200\kms$, and $\eta_{\rm MB} = 1$.  The left and right
panels are for $\lambda_{\rm CE} = 0.1$ and 0.5, respectively.
Hatched regions indicate systems that are persistent X-ray sources
($+45\degr$) or transient ($-45\degr$) (see below).  The solid
histogram that encloses the hatched regions shows the total number of
systems.  The dotted and thick, solid histograms in the right panel are
explained in \S~\ref{sec:obs} and \S~\ref{sec:irr}.

We decide if a system is persistent or transient according to the
standard disk instability model \citep[e.g.,][]{cannizzo82}, wherein
the accretion is transient if the X-ray irradiation temperature at the
disk edge is $T_{\rm irr} \ga 10^4$\,K \citep[see][and references
therein]{vanparadijs96}, a characteristic hydrogen ionization
temperature.  The irradiation temperature at a radius $R$ in the disk
is given by \citep[e.g.,][]{dejong96}
\be
T_{\rm irr}^4 = \frac{L_{\rm X}}{4 \pi \sigma_{\rm SB}R^2}
\frac{H}{R}(1-\gamma)(\xi - 1)~,
\ee
where $\sigma_{\rm SB}$ is the Stefan-Boltzmann constant, $H$ is the
disk scale height, $\gamma$ is the X-ray albedo of the disk, and $\xi
\equiv d\ln H/d\ln R$.  Following, e.g., \citet{king96}, we take the
outer disk radius to be 70\% of the Roche-lobe radius of the NS, and
adopt values of $\xi = 9/7$ \citep{vrtilek90}, $\gamma = 0.9$, and
$H/R = 0.2$ at the disk edge \citep{dejong96}.  Fixing the NS mass and
radius at $1.4\msun$ and 10\,km, respectively, we find the following
expression for the critical mass-transfer rate below which an L/IMXB
is transient:
\be\label{eq:trans}
\frac{\dot{M}_{a,{\rm crit}}}{10^{-10}\mdot} \sim  
(1+q)^{2/3}r_{{\rm L}a}^2 \, 
\left(\frac{P_{\rm orb}}{1\,{\rm hr}}\right)^{4/3}~,
\ee
where $r_{{\rm L}a}$ is the Roche-lobe radius for the NS in units of
the orbital separation.  The value of $\dot{M}_{a,{\rm crit}}$ is not
certain to within a factor of at least a few, and so our numbers and
distributions for transient and persistent systems should be
considered cautiously.

In Fig.~\ref{fig:cedis}, the shapes and extents of the theoretical
distributions for $\lambda_{\rm CE} = 0.1$ and 0.5 are quite similar.
The most notable difference between the two cases is the total number
of systems: $\sim$$10^3$ when $\lambda_{\rm CE} = 0.1$ and
$\sim$$10^5$ when $\lambda_{\rm CE} = 0.5$.  Also prominent in both
panels is the very large number of systems with $P_{\rm orb} \la
2$\,hr, the majority of which are transient according to
eq.~(\ref{eq:trans}).  Note finally that, while IMXBs are much more
favorably produced than LMXBs, very few systems at the current epoch
have donors of mass $>$$1\msun$.  This is simply because the initial
phase of thermal-timescale mass transfer in IMXBs, where a large
fraction of the secondary mass is removed, is relatively short-lived
(see, e.g., Fig.~2a of Paper I).

\subsection{Comparisons with Observation}\label{sec:obs}

Small number statistics, observational selection effects, and sample
incompleteness are all serious issues for the observed LMXBs.  Out of
$\sim$140 Galactic X-ray sources classified as LMXBs with NS
accretors, there are only $\sim$40 systems with measured orbital
periods and a handful with estimated secondary masses.  Usually, the
nondetection of the donor at optical wavelengths is taken to mean that
it is of low mass.  The estimation of X-ray luminosities among the
observed LMXBs that lie outside of globular clusters is complicated by
very poor distance estimates.  Furthermore, for X-ray luminosities
less than $L_{\rm X} \sim 10^{36}\ergs$ ($\dot{M}_a \sim
10^{-10}\mdot$), the observed sample may be quite incomplete.  Even in
light of these problems, some rough comparisons between our
theoretical results and the observations are illuminating.  The most
apparent quantitative discrepancy is in the total numbers of systems.

We define $N_{\rm X}$ to be the total number of L/IMXBs at the current
epoch---both persistent and transient---with secular accretion rates
of $\dot{M}_a > 10^{-10}\mdot$.  Thus, $N_{\rm X}$ quantifies the
number of {\em luminous} X-ray binaries that would be observable over
a large fraction of the Galactic volume. Table~1 lists values of $\log
N_{\rm X}$ for different parameter sets, from which we see that
$N_{\rm X}$ ranges from $\sim$3 to $>$400 times the the total of
$\sim$140 systems observed in the Galaxy.  We elaborate in the next
section on this rather severe overproduction problem.

In Fig.~\ref{fig:cedis}, for $\lambda_{\rm CE} = 0.5$, we have
overlayed (dotted histogram) on the middle panel the distribution of
37 measured orbital periods for LMXBs outside of globular clusters
\citet{liu01}, which has been multiplied by a factor of 3000 for
comparison purposes.  With such small numbers (1--9) per bin,
meaningful comparisons with our theoretical distribution are
difficult.  It may be that we are underproducing systems with $P_{\rm
orb}\ga 10\day$ and overproducing short-period LMXBs with $P_{\rm
orb}\la 0.1\day$. Our simulations indicate that most of the
short-period systems are transient, which may aid in explaining a
possible discrepancy.  Furthermore, binaries with $P_{\rm orb}\la
2$\,hr are driven by GR with characteristically low X-ray luminosities
of $L_{\rm X} < 10^{35}$--$10^{36}\ergs$, making them more difficult
to discover.  We compute the fraction, $f_{\rm SP}$, of short-period
($P_{\rm orb} < 2$\,hr) LMXBs at the current epoch, and find that
typically $f_{\rm SP} \sim 0.3$ (Table~1), which in itself is not in
serious conflict with observations, for reasons already mentioned.

In our simulations, we actually underestimate the number of
short-period ($P_{\rm orb}\la 2$\,hr) LMXBs at the current epoch.
These binaries evolve from systems with initial orbital periods below
the bifurcation period of $\sim$18\,hr (see Paper I and references
therein), and reach minimum periods of $\sim$10\,min to $\sim$1.5\,hr.
However, for technical reasons, our X-ray binary calculations are
terminated not long after the minimum period is reached.  The
subsequent evolution is driven by GR at low rates, but may last for
billions of years.  This makes the discrepancy with the observations
somewhat worse.  On the other hand, we note that 3 ultracompact binary
systems have been discovered within the past year alone, so the
discovery probability for these systems may be increasing.

Extrapolating the data available for 16 Galactic LMXBs monitored by
{\em RXTE}/ASM, \citet{grimm02} have attempted to compute a cumulative
X-ray luminosity distribution, corrected for the fraction of the
Galactic volume observable by the ASM.  For luminosities of
$\ga$$10^{36}\ergs$, we expect that the observed sample should be
reasonably complete.  As a point of comparison, we have chosen to
compute the ratio of the number of LMXBs with X-ray luminosities of
$L_{\rm X} = 10^{36}$--$10^{37}\ergs$ to the number with
$>$$10^{37}\ergs$.  The results of \citet{grimm02} indicate that ratio
is $\sim$1.5.  Our theoretical cumulative distribution of mass
accretion rates is shown as the thin line in Fig.~\ref{fig:mdotcum};
the thick line will be discussed in \S~\ref{sec:irr}.  For the two
luminosity ranges given above, we find a larger number ratio of
$\sim$5.3.  It is not clear if this particular discrepancy between our
theoretical results and the observations is especially significant.


\section{DISCUSSION}\label{sec:con}

In this section, we devote a short discussion to each of three
important topics that relate to L/IMXB evolution.  Binary millisecond
pulsars are discussed, with emphasis on the orbital period
distribution and the long-standing birthrate problem.  We also address
the possible importance of X-ray irradiation of the donor stars in
L/IMXB.  Finally, we briefly investigate the possibility of forming
low-mass black holes in L/IMXBs via accretion-induced collapse.

\subsection{Binary Millisecond Radio Pulsars \\ and the Birthrate Problem}

Binary millisecond radio pulsars (BMPs) are widely thought to be the
evolutionary descendants of L/IMXBs \citep{alpar82,joss83}.  Using our
BPS code and library of evolutionary sequences, we compute the orbital
period distribution of BMPs in the following way.  Most sequences are
terminated after the He or HeCO white-dwarf core of the donor star is
exposed (see Paper I).  For the binaries that contract to $P_{\rm orb}
\la 2$\,hr, the evolution ends not long after the period minimum is
reached; as mentioned above, further evolution to longer periods is
not followed.  At the end of each sequence, we know the orbital period
and mass of the secondary.  Each sequence $i$ and the corresponding
final orbital period are weighted by the formation efficiency
$\mathcal{F}_i$, and the results are accumulated to generate a
histogram.  We have not attempted to include any estimates for the
pulsar lifetime or detectability in our analysis.

\begin{figure*}
\begin{minipage}[t]{0.5\linewidth}
\centerline{\epsfig{file=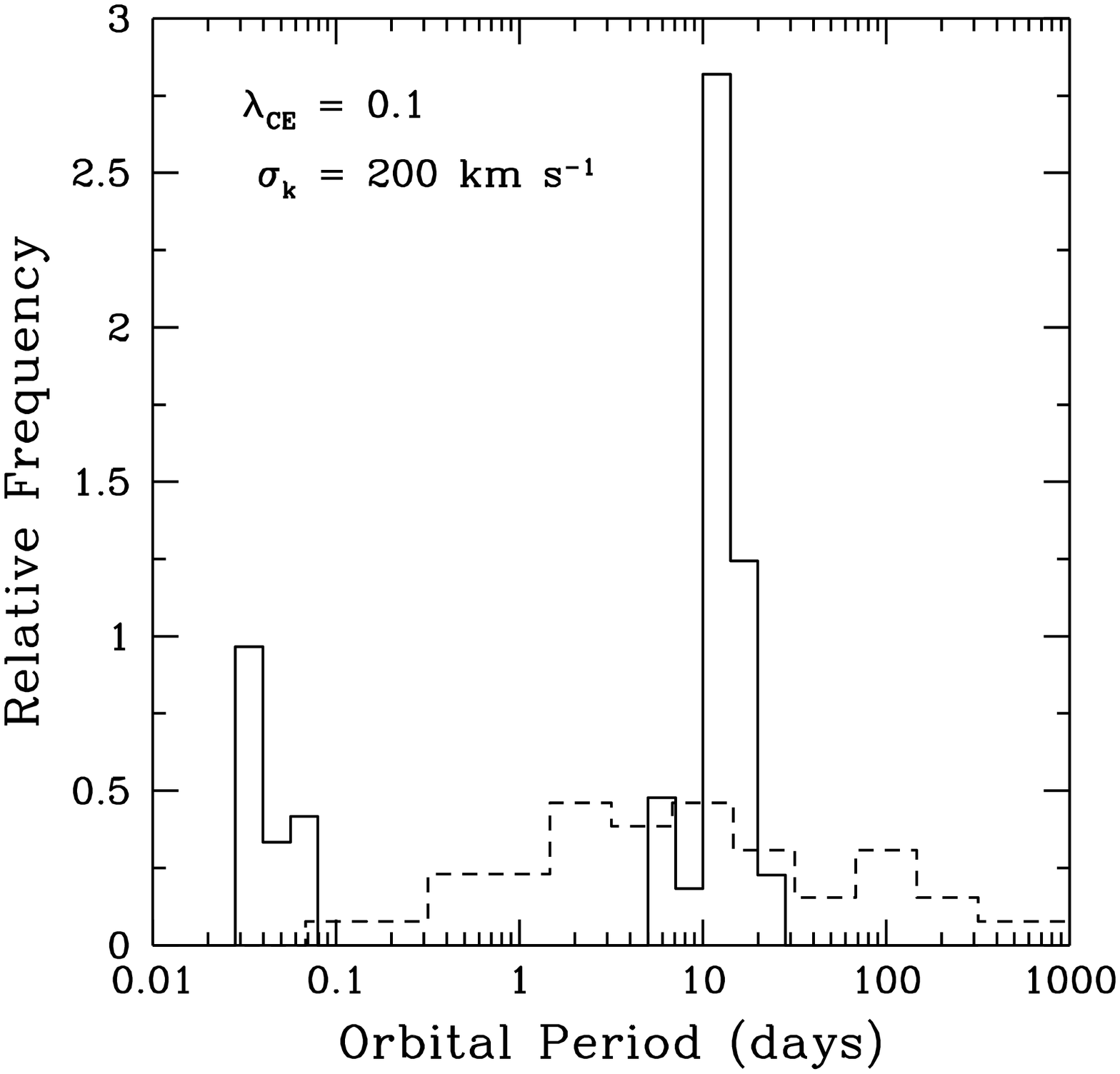,width=0.9\linewidth}}
\end{minipage}
\hfill
\begin{minipage}[t]{0.5\linewidth}
\centerline{\epsfig{file=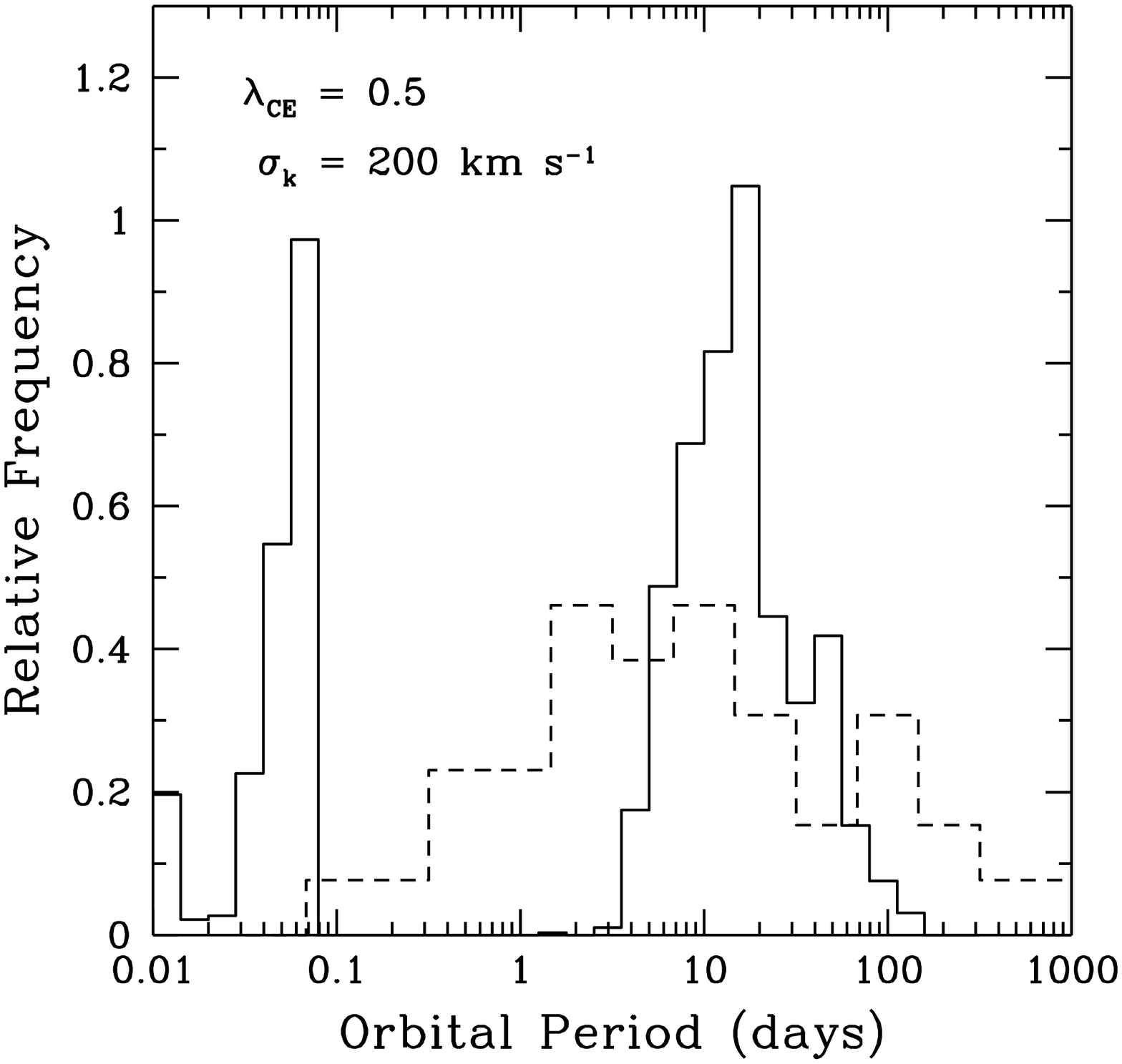,width=0.9\linewidth}}
\end{minipage}
\caption{Orbital period distribution of binary millisecond radio
pulsars.  The solid and dashed histograms are for the theoretical and
observed systems, respectively.  Both distribution are normalized to
unit area.
\label{fig:bmsp}}
\end{figure*}

Two examples of the calculated orbital period distribution are shown
in Fig.~\ref{fig:bmsp} (solid histogram), where $y= 0.0$, $\eta_{\rm
MB} = 1.0$, $\sigma_k = 200\kms$, and $\lambda_{\rm CE} = 0.1$ and
0.5. Systems in Fig.~\ref{fig:bmsp} with $P_{\rm orb} \la 2$\,hr are
meant only to indicate the relative proportions of short-period and
long-period BMPs produced in our simulations.  Overlayed is the
period distribution of the observed systems \citep[dashed histogram;
taken from][]{taam00}.  Both distributions are normalized to unit
area.  Clearly, neither parameter set adequately reproduces the
observed distribution, and the agreement is extremely poor for
$\lambda_{\rm CE} = 0.1$.  Larger values of $\lambda_{\rm CE}$ are
strongly favored, since L/IMXBs form over a much wider range of
initial periods and donor masses (see Fig.~\ref{fig:incip}, top
panels) than when $\lambda_{\rm CE} \sim 0.1$, which results in a
wider range of BMP orbital periods.  The same conclusion was reached
by \citet{willems02}, who used simplified models of L/IMXB evolution
in a focused population synthesis study of BMPs.  However,
$\lambda_{\rm CE}\sim 0.5$ is problematic, since the number of
luminous L/IMXBs at the current epoch is greatly overproduced, by
factors of $\ga$100, relative to the observed number, as noted in the
last section.  This problem emerges again when one considers the
birthrates of BMPs based on the observed sample.

\citet{lorimer95} and \citep{cordes97} each conservatively estimate
the total Galactic birthrate of BMPs to be $\sim$$10^{-6}\yr^{-1}$.
Their likelihood analyses considered $\sim$20 BMPs and included known
pulsar selection effects, a model for the spatial distribution of
BMPs, and estimated distance errors.  \citet{lorimer95} notes that
several uncertainties may increase the birthrate by a factor of 10, to
$\sim$$10^{-5}\yr^{-1}$.  In addition, the pulsar spin-down ages used
in these studies may systematically overestimate the true ages of the
systems \citep[e.g.,][]{lorimer95,hansen98}, leading to larger actual
birthrates. Given the $\sim$100 observed LMXBs in the Galaxy and a
typical observable X-ray lifetime of $\langle \tau_{\rm X} \rangle
\sim 10^9\yr$, the the semi-empirical birthrate of LMXBs is
$\sim$$10^{-7}\yr^{-1}$, which is 10--100 times lower than the BMP
birthrate.  \citet{kulkarni88} were the first to point out this
potentially important discrepancy.

Recall from \S~\ref{sec:incip} that the birthrates of L/IMXBs lie in
the range $\sim$$10^{-6}$--$10^{-4}\yr^{-1}$, which covers the range
of semi-empirical BMP birthrates.  The largest L/IMXB birthrates,
which may be required if the BMP birthrate is $\sim$$10^{-5}\yr^{-1}$,
result from $\lambda_{\rm CE} \sim 0.3$--0.5, which, in turn, yields
far too many luminous X-ray sources at the current epoch.  Although
the active X-ray lifetime is not relevant for our theoretical L/IMXB
birthrate calculations, this is an important quantity to calculate.

We define the mean {\em luminous} X-ray lifetime from our simulations
to be
\be\label{eq:xray}
\langle \tau_{\rm X} \rangle = \sum \mathcal{F}_i \Delta t_i/ 
\sum \mathcal{F}_i~,
\ee
where $\Delta t_i$ is the total time spent by sequence $i$ in the
interval $\dot{M}_a > 10^{-10}\mdot$, and the sums are over all
sequences with $M_d < 4$ selected in the population synthesis
calculation.  We have not attempted to correct $\langle \tau_{\rm
X}\rangle$ for systems that are transient according to the disk
instability model, since the transient duty cycle is unknown.  From
Table~1 we see that $\langle \tau_{\rm X}\rangle$ is consistently
$\sim$$5\times 10^8\yr$.  This justifies the past use of $\langle
\tau_{\rm X}\rangle\sim 10^9\yr$ as a typical LMXB lifetime.  However,
it is essentially this long X-ray lifetime that yields such large
numbers of luminous sources at the current epoch in our model
calculations.  We now discuss a possible resolution to this serious
conflict.

\subsection{Irradiation-Induced Mass-Transfer Cycles}\label{sec:irr}

Our binary evolution calculations do not account for X-ray irradiation
effects on the secondary, which can dramatically change the evolution
of the system, either by driving winds \citep{ruderman89} or the
expansion of the star \citep{podsi91}.  Cyclic mass transfer may
result from the irradiation, characterized by short episodes of
enhanced mass transfer and long detached phases
\citep[e.g.,][]{hameury93,harpaz94}. Such cycles may then
significantly reduce the total X-ray active lifetime, and thus resolve
the L/IMXB overproduction problem and the discrepancy between the
semi-empirical birthrates of BMPs and LMXBs.

\begin{inlinefigure}
\centerline{\epsfig{file=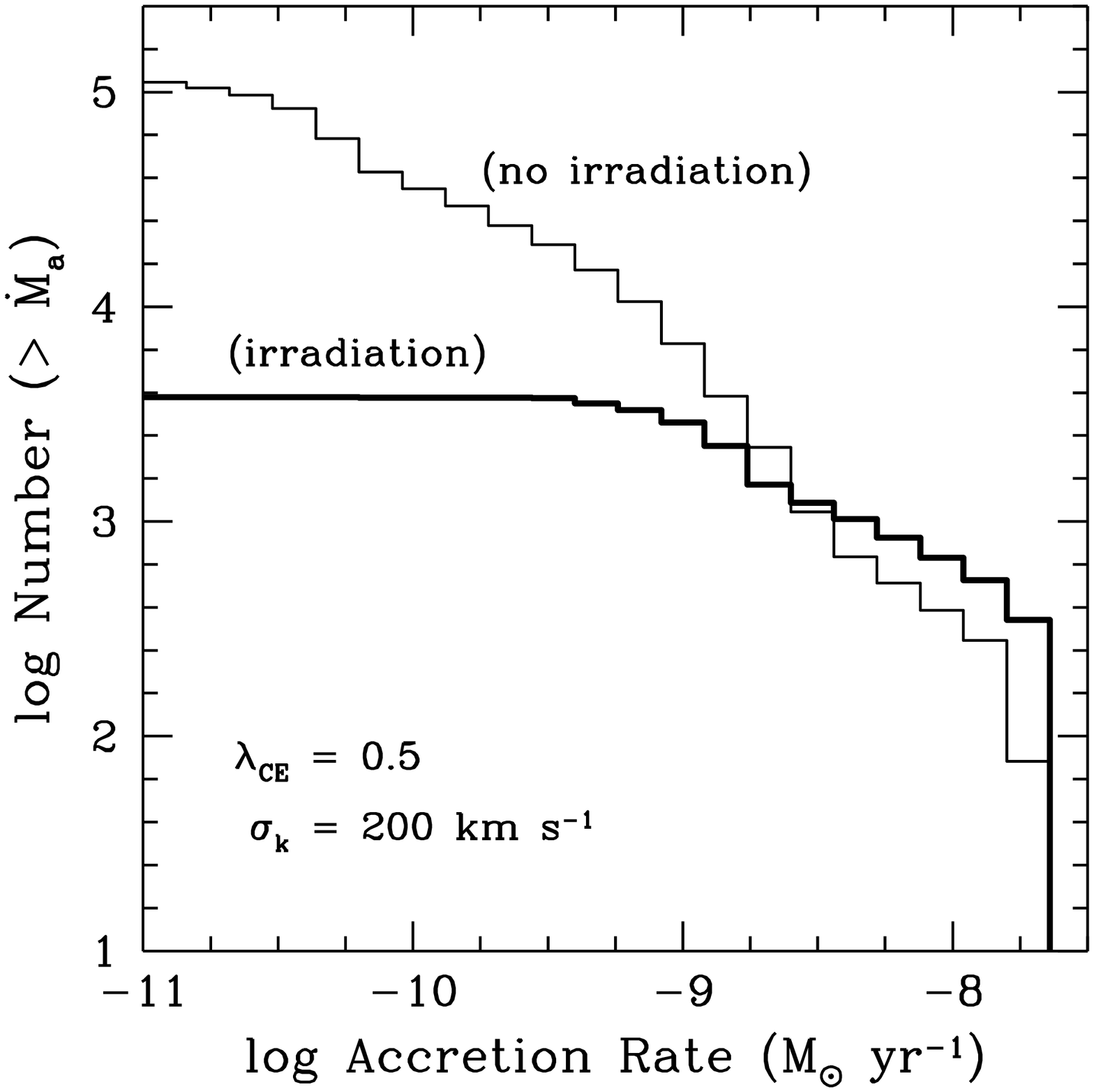,width =
\linewidth}}
\caption{Cumulative distribution of accretion rates at the current
epoch for the same systems used to generate Fig.~\ref{fig:cedis}
($\lambda_{\rm CE} = 0.5$).  Here we include both persistent and
transient systems.  The thick line corresponds to our ad hoc inclusion
of irradiation effects.
\label{fig:mdotcum}}
\end{inlinefigure}

X-ray irradiation affects stars of mass $\la$$1.5\msun$ by ionizing
the hydrogen at the base of the irradiated surface layer and
disrupting the surface convection zone.  By changing the surface
boundary condition in this way, the star would like to expand to a new
thermal-equilibrium radius \citep{podsi91}, with more dramatic
expansion as the total stellar mass is decreased and the fractional
mass in the surface convection zone is increased.  Intermediate-mass
stars in IMXBs, which are initially fully radiative, may be strongly
affected by irradiation once the mass is reduced to $\la$1--$1.5\msun$
and a surface convection zone appears, after the early, rapid phase of
thermal-timescale mass transfer.

An irradiation temperature of $\ga$$10^4$\,K---a characteristic
ionization temperature for hydrogen---gives a critical X-ray flux of
$S_c \sim 10^{11}$--$10^{12}\flux$, above which irradiation is
important.  Letting $S_{c,11}$ be the critical flux in units of
$10^{11}\flux$, and $\dot{M}_{a,-8}$ be the accretion rate onto the NS
in units of $10^{-8}\mdot$, we estimate the maximum orbital period for
which irradiation is important:
\be
P_{\rm orb} \sim 70\,{\rm day} \,
\left(\frac{\epsilon\,\dot{M}_{a,-8}}{S_{c,11}}\right)^{3/4}~, 
\ee
where $\epsilon < 1$ is a factor that takes into account the geometry
of the accretion disk and star, albedo of the star, and fraction of
X-rays that penetrate below the stellar photosphere \citep{hameury93}.
The value of $\epsilon$ is uncertain, as is the effect of irradiation
on stellar and binary stellar evolution.  However, it is at least
quite plausible, and perhaps likely, that X-ray irradiation
significantly alters the evolution of most L/IMXBs.

When irradiation effects are relatively moderate, the results of
\citet{hameury93} indicate that irradiation-induced mass-transfer
cycles do not typically change the secular evolution of $P_{\rm orb}$
or $M_d$.  Under these circumstances, the inclusion of irradiation
would not change the $P_{\rm orb}$ and $M_d$ distributions in
Fig.~\ref{fig:cedis}.  However, the distribution of $\dot{M}_a$ would
change significantly.  Suppose that the donors in all L/IMXBs are
affected by irradiation in the same way, such that the mass-transfer
rate is on average enhanced by a factor of $f_{\rm en} >1$ during
episodes of Roche-lobe overflow.  We then expect the $\log \dot{M}_a$
distribution in Fig.~\ref{fig:cedis} to be shifted to the right by an
amount $\log f_{\rm en}$, but still cut off at the Eddington limit.
Furthermore, the distribution is everywhere decreased in height by an
amount $-\log f_{\rm en}$, since the X-ray lifetime for each system is
reduced by a factor $f^{-1}$.  In order to alleviate the L/IMXB
overproduction problem and the BMP birthrate problem, we perhaps need
to have $\langle \tau_{\rm X} \rangle$ reduced by a factor of
$\ga$100.  This would require $f_{\rm en} \ga 100$, depending on the
shape of the $\dot{M}_a$ distribution calculated {\em without}
irradiation.

In Figs.~\ref{fig:cedis} ($\lambda_{\rm CE} = 0.5$) and
\ref{fig:mdotcum} we have overlayed (thick, solid histogram) a
distribution that may very roughly illustrate the net effects of
irradiation for an illustrative enhancement factor of $f_{\rm en} =
30$. To get this result, we simply multiplied the mass-transfer rate
from the donor by $f_{\rm en}$ and all times by $f_{\rm en}^{-1}$ in
each of the selected evolutionary sequences, regardless of the
instantaneous values of $M_d$ or $P_{\rm orb}$.  We have chosen
$f_{\rm en} = 30$ since $\langle \tau_{\rm X} \rangle$ and $N_{\rm X}$
are reduced by a factor of 10.  The resulting $\dot{M}_a$ distribution
(Fig.~\ref{fig:cedis}) is confined to the range $\dot{M}_a >
10^{-10}\mdot$, and we find that most of these systems are persistent.
From the corresponding cumulative distribution shown in
Fig.~\ref{fig:mdotcum}, we find that the ratio of the number of LMXBs
with luminosities of $10^{36}$--$10^{37}\ergs$ to the number with
$L_{\rm X} > 10^{37}\ergs$ is $\sim$0.6, smaller than the observed
value (see \S~\ref{sec:obs}).  However, irradiation effects at least
produce the desired outcome of reducing this ratio from our earlier
quoted theoretical value of $\sim$5.3.  Of course, our treatment of
irradiation effects is greatly oversimplified, and much more detailed
calculations are requited.



\subsection{Low-Mass Black Holes}

It is generally assumed that the NS cannot accrete at rates exceeding
the Eddington limit ($\dot{M}_{\rm Edd} \sim 10^{-8}\mdot$).  Material
that is donated faster than $\dot{M}_{\rm Edd}$, such as occurs during
thermal-timescale mass transfer, will likely be ejected from the
system, possibly as a radiatively driven wind from the accretion disk
or in the form of relativistic jets.  Evidence for both of these
processes is seen in the X-ray binary SS 433 \citep{blundell01}, a
system known to be in a phase of super-Eddington mass transfer.

Our library of evolutionary sequences were computed with one
reasonable, though heuristic, prescription for the mass capture
fraction $\beta$:
\be\label{ch3_eq:beta}
\beta = 
\begin{cases}
b~, & ~~|\dot{M}_2| < \dot{M}_{\rm Edd} \\
b\,\dot{M}_{\rm Edd}/|\dot{M}_2|~, & ~~|\dot{M}_2| > \dot{M}_{\rm Edd}~,
\end{cases}
\ee
where $b \le 1$ is a constant.  This formula limits the accretion rate
to be $\dot{M}_a < b\,\dot{M}_{\rm Edd}$.  We used a specific value of
$b = 0.5$ in our calculations.  

The distribution of final NS masses (Fig.~\ref{fig:nsdis}) was
computed in the same way as the orbital-period distribution for BMPs.
Masses up to $\sim$$2.5\msun$ were reached in these simulations.  Most
modern NS equations of state give maximum masses of $\sim$2--$3\msun$.
Thus, even for $b = 0.5$, it is possible that a significant fraction
of L/IMXBs ultimately contain a low-mass black hole.  If we had used
$b = 1$, our NS mass distribution would have broadened to include
masses up to $\sim$$3.6\msun$, and perhaps several tens of percent of
NSs in L/IMXBs would collapse to black holes.  No binary that contains
a compact object of mass $\sim$2--$3\msun$ has yet been confirmed
observationally.  Such systems would not exhibit X-ray bursts and are
unlikely to show the twin kHz quasi-periodic oscillations seen in
LMXBs \citep{vanderklis00}.  The inclusion of X-ray irradiation on the
donor star, and the possible dramatic decrease in the X-ray lifetime,
may generally yield final NS masses much closer to the initial mass of
$1.4\msun$.  Perhaps this could explain why the measured NS masses in
BMPs lie in a narrow range about $\sim$$1.4\msun$ \citep{thorsett99}.

\begin{inlinefigure}
\centerline{\epsfig{file=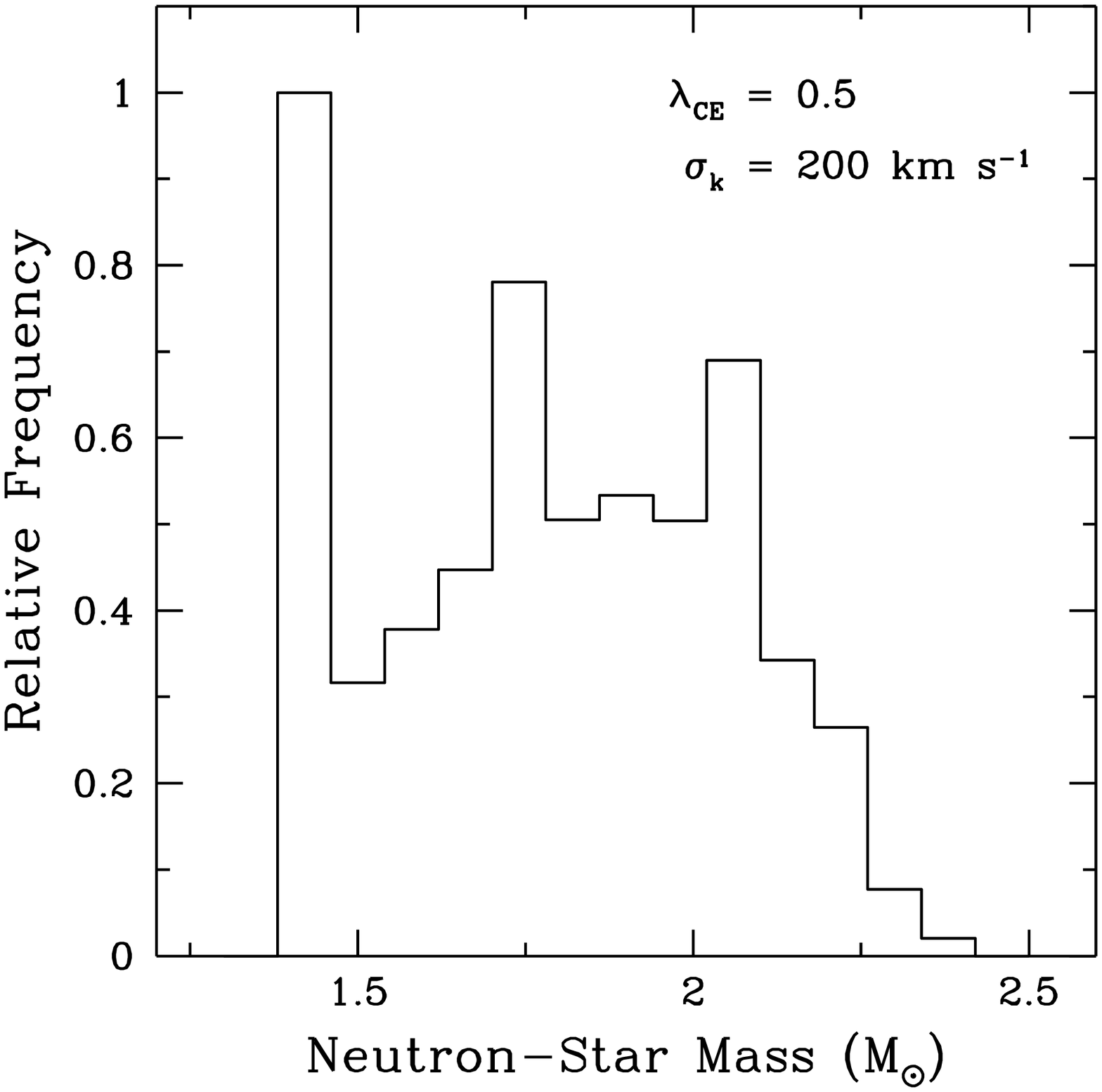,width = \linewidth}}
\caption{Distribution of final NS masses for BPS parameters $y = 0.0$,
$\eta_{\rm MB} = 1$, $\sigma_k = 200\kms$, and $\lambda_{\rm CE} =
0.5$.
\label{fig:nsdis}}
\end{inlinefigure}


\section{Summary and Outlook}

Here we list the key points and results of this paper, in decreasing
order of importance.

\medskip

1. This is the first population synthesis study of L/IMXBs that
incorporates detailed evolutionary calculations.  With this addition,
we are able to follow a population of L/IMXBs from (i) the incipient
stage, (ii) to the current epoch, and finally (iii) to the remnant
state when they presumably become BMPs.  We are thus able to
meaningfully compare our results with the sample of observed LMXBs and
BMPs.

2. We have demonstrated that incipient IMXBs outnumber incipient LMXBs
typically by a factor of $\ga$5 (see Table~1).  Since IMXBs may evolve
to resemble observed LMXBs, we claim that the majority of observed
systems may have started their lives with intermediate-mass donor
stars.

3. We find that rather large values of $\lambda_{\rm CE}$ ($\sim$0.5)
are required in order (i) for the theoretical BMP $P_{\rm orb}$
distribution to even remotely resemble the observed distribution, and
(ii) to yield L/IMXB birthrates which are consistent with the
semi-empirical BMP birthrates.  However, we have discovered that such
values of $\lambda_{\rm CE}$ lead to a dramatic overproduction of the
number of luminous X-ray binaries in the Galaxy at the current epoch,
by factors of $\sim$100--1000.

4. The overproduction problem and the discrepancy between
semi-empirical BMP and LMXB birthrates may be resolved if the mean
X-ray lifetime of L/IMXBs is reduced by a factor of $\ga$100.  Cyclic
mass transfer, induced by the X-ray irradiation of the donor star, may
have the desired outcome.  It is plausible X-ray irradiation strongly
affects the evolution of most L/IMXBs.

5. Eddington-limited accretion onto the NS can lead to large NS masses
of $\sim$2--$4\msun$.  It is then possible that a significant fraction
of NSs in L/IMXBs collapse to low-mass black holes.  To our knowledge,
the only way to confirm the presence of a low-mass black hole in an
observed LMXB is to measure the mass dynamically, which is extremely
difficult in general.  If the X-ray lifetimes of L/IMXBs are reduced
substantially, by, e.g., the effects of X-ray irradiation, so too is
the amount of mass that NSs may accrete.  This may also explain why
the measured NS masses in observed BMPs lie near $\sim$$1.4\msun$.

\bigskip

We conclude by listing a number of ways that our work may be extended
and improved.

\medskip

1.  Since X-ray irradiation of the donor star may be an extremely
important component of L/IMXB evolution, it is important to have a
better quantitative understanding of this process.  The problem is
inherently three-dimensional, and must be treated as such in order to
obtain meaningful quantitative results.  Important first steps in this
regard have been made by \citet{beer02}, but much work remains to be
done.

2. A detailed comparison of theoretical models of L/IMXBs formation
and evolution with the observed population requires that we compute
the spatial trajectories of the synthesized systems in a realistic
Galactic gravitational potential.  It then becomes possible to
generate theoretical X-ray flux distributions as well as distributions
in Galactic latitude and longitude.  A treatment of observational
selection effects, such as instrumental flux limits and X-ray
absorption, also becomes possible.
    
3. Our library of L/IMXB evolutionary sequences will soon be
incorporated into a sophisticated dynamical Monte Carlo code to study
the evolution of globular clusters \citep[][and references
  therein]{fregeau03}.  The code now incorporates direct numerical
integrations of single and binary-binary dynamical interactions, a
mass spectrum of stars, and analytic treatments of single-star
evolution.  The inclusion of our library of X-ray binary calculations
will make it possible to study directly the LMXB and BMP populations
in globular clusters.

4. An appropriately detailed treatment of the tidal evolution of
incipient L/IMXBs immediately after the SN should be included in
future studies.  Orbital circularization and spin synchronization of
the secondary, coupled with orbital angular momentum loss due to GR,
as well the loss of spin angular momentum that results from MB should
be considered.

5.  It would be extremely advantageous if L/IMXB evolutionary
calculations could be carried out at least 100 times faster than is
possible at present.  On a reasonably fast workstation, a typical
computing time is currently $\sim$20\,min, so that, realistically,
several days of computing time would be required to regenerate our
current library of sequences.  The development of an ultrafast
Henyey-type stellar evolution code, along with increased processor
speed over the next several years may make it possible to carry out
1000 L/IMXB evolutions in $\sim$10\,hr of computing time.



\acknowledgements

EP was supported by NASA and the Chandra Postdoctoral Fellowship
program through grant number PF2-30024, awarded by the Chandra X-ray
Center, which is operated by the Smithsonian Astrophysical Observatory
for NASA under contract NAS8-39073.



\begin{thebibliography}{68}
\expandafter\ifx\csname natexlab\endcsname\relax\def\natexlab#1{#1}\fi

\bibitem[{{Abt} \& {Levy}(1978)}]{abt78}
{Abt}, H.~A. \& {Levy}, S.~G. 1978, \apjs, 36, 241

\bibitem[{{Alpar} {et~al.}(1982){Alpar}, {Cheng}, {Ruderman}, \&
  {Shaham}}]{alpar82}
{Alpar}, M.~A., {Cheng}, A.~F., {Ruderman}, M.~A., \& {Shaham}, J. 1982, \nat,
  300, 728

\bibitem[{{Arzoumanian} {et~al.}(2002){Arzoumanian}, {Chernoff}, \&
  {Cordes}}]{arzoumanian02}
{Arzoumanian}, Z., {Chernoff}, D.~F., \& {Cordes}, J.~M. 2002, \apj, 568, 289

\bibitem[{{Beer} \& {Podsiadlowski}(2002)}]{beer02}
{Beer}, M.~E. \& {Podsiadlowski}, P. 2002, \mnras, 335, 358

\bibitem[{{Belczynski} {et~al.}(2002){Belczynski}, {Kalogera}, \&
  {Bulik}}]{belczynski02}
{Belczynski}, K., {Kalogera}, V., \& {Bulik}, T. 2002, \apj, 572, 407

\bibitem[{{Blaauw}(1961)}]{blaauw61}
{Blaauw}, A. 1961, \bain, 15, 265+

\bibitem[{{Blundell} {et~al.}(2001){Blundell}, {Mioduszewski}, {Muxlow},
  {Podsiadlowski}, \& {Rupen}}]{blundell01}
{Blundell}, K.~M., {Mioduszewski}, A.~J., {Muxlow}, T.~W.~B., {Podsiadlowski},
  P., \& {Rupen}, M.~P. 2001, \apjl, 562, L79

\bibitem[{{Boersma}(1961)}]{boersma61}
{Boersma}, J. 1961, \bain, 15, 291

\bibitem[{{Brown} {et~al.}(2001){Brown}, {Heger}, {Langer}, {Lee}, {Wellstein},
  \& {Bethe}}]{brown01}
{Brown}, G.~E., {Heger}, A., {Langer}, N., {Lee}, C.-H., {Wellstein}, S., \&
  {Bethe}, H.~A. 2001, New Astronomy, 6, 457

\bibitem[{{Cannizzo} {et~al.}(1982){Cannizzo}, {Ghosh}, \&
  {Wheeler}}]{cannizzo82}
{Cannizzo}, J.~K., {Ghosh}, P., \& {Wheeler}, J.~C. 1982, \apjl, 260, L83

\bibitem[{{Cappellaro} {et~al.}(1999){Cappellaro}, {Evans}, \&
  {Turatto}}]{cappellaro99}
{Cappellaro}, E., {Evans}, R., \& {Turatto}, M. 1999, \aap, 351, 459

\bibitem[{{Chakrabarty} \& {Morgan}(1998)}]{chakrabarty98}
{Chakrabarty}, D. \& {Morgan}, E.~H. 1998, \nat, 394, 346

\bibitem[{{Cordes} \& {Chernoff}(1997)}]{cordes97}
{Cordes}, J.~M. \& {Chernoff}, D.~F. 1997, \apj, 482, 971

\bibitem[{{De Greve} \& {De Loore}(1977)}]{degreve77}
{De Greve}, J.-P. \& {De Loore}, C. 1977, \apss, 50, 75

\bibitem[{{de Jong} {et~al.}(1996){de Jong}, {van Paradijs}, \&
  {Augusteijn}}]{dejong96}
{de Jong}, J.~A., {van Paradijs}, J., \& {Augusteijn}, T. 1996, \aap, 314, 484

\bibitem[{{Delgado} \& {Thomas}(1981)}]{delgado81}
{Delgado}, A.~J. \& {Thomas}, H.-C. 1981, \aap, 96, 142

\bibitem[{{Dewi} {et~al.}(2002){Dewi}, {Pols}, {Savonije}, \& {van den
  Heuvel}}]{dewi02}
{Dewi}, J.~D.~M., {Pols}, O.~R., {Savonije}, G.~J., \& {van den Heuvel},
  E.~P.~J. 2002, \mnras, 331, 1027

\bibitem[{{Dewi} \& {Tauris}(2000)}]{dewi00}
{Dewi}, J.~D.~M. \& {Tauris}, T.~M. 2000, \aap, 360, 1043

\bibitem[{{Dewi} \& {Tauris}(2001)}]{dewi01}
{Dewi}, J.~D.~M. \& {Tauris}, T.~M. 2001, in ASP Conf. Ser. 229: Evolution of
  Binary and Multiple Star Systems, 255

\bibitem[{{Fregeau} {et~al.}(2003){Fregeau}, {Gurkan}, {Joshi}, \&
{Rasio}}]{fregeau03} {Fregeau}, J.~M., {Gurkan}, M.~A., {Joshi},
K.~J., \& {Rasio}, F.~A. 2003, \apj, submitted, astro-ph/0301521

\bibitem[{{Garmany} {et~al.}(1980){Garmany}, {Conti}, \& {Massey}}]{garmany80}
{Garmany}, C.~D., {Conti}, P.~S., \& {Massey}, P. 1980, \apj, 242, 1063

\bibitem[{{Grimm} {et~al.}(2002){Grimm}, {Gilfanov}, \& {Sunyaev}}]{grimm02}
{Grimm}, H.-J., {Gilfanov}, M., \& {Sunyaev}, R. 2002, \aap, 391, 923

\bibitem[{{Habets}(1986{\natexlab{a}})}]{habets86a}
{Habets}, G.~M.~H.~J. 1986{\natexlab{a}}, \aap, 165, 95

\bibitem[{{Habets}(1986{\natexlab{b}})}]{habets86b}
---. 1986{\natexlab{b}}, \aap, 167, 61

\bibitem[{{Hameury} {et~al.}(1993){Hameury}, {King}, {Lasota}, \&
  {Raison}}]{hameury93}
{Hameury}, J.~M., {King}, A.~R., {Lasota}, J.~P., \& {Raison}, F. 1993, \aap,
  277, 81

\bibitem[{{Hansen} \& {Phinney}(1997)}]{hansen97}
{Hansen}, B.~M.~S. \& {Phinney}, E.~S. 1997, \mnras, 291, 569

\bibitem[{{Hansen} \& {Phinney}(1998)}]{hansen98}
---. 1998, \mnras, 294, 569

\bibitem[{{Harpaz} \& {Rappaport}(1994)}]{harpaz94}
{Harpaz}, A. \& {Rappaport}, S. 1994, \apj, 434, 283

\bibitem[{{Hurley} {et~al.}(2000){Hurley}, {Pols}, \& {Tout}}]{hurley00}
{Hurley}, J.~R., {Pols}, O.~R., \& {Tout}, C.~A. 2000, \mnras, 315, 543

\bibitem[{{Ivanova} {et~al.}(2003){Ivanova}, {Belczynski}, {Kalogera},
{Rasio}, \& {Taam}}]{ivanova03} {Ivanova}, N., {Belczynski}, K.,
{Kalogera}, V., {Rasio}, F.~A., \& {Taam}, R.~E. 2003, \apj,
submitted, astro-ph/0210267

\bibitem[{{Joss} \& {Rappaport}(1983)}]{joss83}
{Joss}, P.~C. \& {Rappaport}, S.~A. 1983, \nat, 304, 419

\bibitem[{{Kalogera}(1998)}]{kalogera98b}
{Kalogera}, V. 1998, \apj, 493, 368

\bibitem[{{Kalogera} \& {Webbink}(1996)}]{kalogera96a}
{Kalogera}, V. \& {Webbink}, R.~F. 1996, \apj, 458, 301

\bibitem[{{Kalogera} \& {Webbink}(1998)}]{kalogera98a}
---. 1998, \apj, 493, 351

\bibitem[{{King} {et~al.}(1996){King}, {Kolb}, \& {Burderi}}]{king96}
{King}, A.~R., {Kolb}, U., \& {Burderi}, L. 1996, \apjl, 464, L127+

\bibitem[{{King} \& {Ritter}(1999)}]{king99}
{King}, A.~R. \& {Ritter}, H. 1999, \mnras, 309, 253

\bibitem[{{Kippenhahn} \& {Weigert}(1967)}]{kippenhahn67}
{Kippenhahn}, R. \& {Weigert}, A. 1967, Zeitschrift Astrophysics, 65, 251

\bibitem[{{Kolb} {et~al.}(2000){Kolb}, {Davies}, {King}, \& {Ritter}}]{kolb00}
{Kolb}, U., {Davies}, M.~B., {King}, A., \& {Ritter}, H. 2000, \mnras, 317, 438

\bibitem[{{Kroupa} {et~al.}(1993){Kroupa}, {Tout}, \& {Gilmore}}]{kroupa93}
{Kroupa}, P., {Tout}, C.~A., \& {Gilmore}, G. 1993, \mnras, 262, 545

\bibitem[{{Kulkarni} \& {Narayan}(1988)}]{kulkarni88}
{Kulkarni}, S.~R. \& {Narayan}, R. 1988, \apj, 335, 755

\bibitem[{{Langer} \& {Maeder}(1995)}]{langer95}
{Langer}, N. \& {Maeder}, A. 1995, \aap, 295, 685

\bibitem[{{Liu} {et~al.}(2001){Liu}, {van Paradijs}, \& {van den
  Heuvel}}]{liu01}
{Liu}, Q.~Z., {van Paradijs}, J., \& {van den Heuvel}, E.~P.~J. 2001, \aap,
  368, 1021

\bibitem[{{Lorimer}(1995)}]{lorimer95}
{Lorimer}, D.~R. 1995, \mnras, 274, 300

\bibitem[{{Miller} \& {Scalo}(1979)}]{miller79}
{Miller}, G.~E. \& {Scalo}, J.~M. 1979, \apjs, 41, 513

\bibitem[{{Nelson} {et~al.}(1986){Nelson}, {Rappaport}, \& {Joss}}]{nelson86}
{Nelson}, L.~A., {Rappaport}, S.~A., \& {Joss}, P.~C. 1986, \apj, 304, 231

\bibitem[{{Pfahl} {et~al.}(2002){Pfahl}, {Rappaport}, \&
  {Podsiadlowski}}]{pfahl02c}
{Pfahl}, E., {Rappaport}, S., \& {Podsiadlowski}, P. 2002, \apj, 573, 283

\bibitem[{{Podsiadlowski}(1991)}]{podsi91}
{Podsiadlowski}, P. 1991, \nat, 350, 136

\bibitem[{{Podsiadlowski} {et~al.}(1992){Podsiadlowski}, {Joss}, \&
  {Hsu}}]{podsi92}
{Podsiadlowski}, P., {Joss}, P.~C., \& {Hsu}, J.~J.~L. 1992, \apj, 391, 246

\bibitem[{{Podsiadlowski} \& {Rappaport}(2000)}]{podsi00}
{Podsiadlowski}, P. \& {Rappaport}, S. 2000, \apj, 529, 946

\bibitem[{{Podsiadlowski} {et~al.}(2002){Podsiadlowski}, {Rappaport}, \&
  {Pfahl}}]{podsi02}
{Podsiadlowski}, P., {Rappaport}, S., \& {Pfahl}, E.~D. 2002, \apj, 565, 1107

\bibitem[{{Podsiadlowski} {et~al.}(2003){Podsiadlowski}, {Rappaport},
\& {Han}}]{podsi03} {Podsiadlowski}, P., {Rappaport}, S., \& {Han},
Z. 2003, \mnras, accepted, astro-ph/0207153

\bibitem[{{Pols}(1994)}]{pols94}
{Pols}, O.~R. 1994, \aap, 290, 119

\bibitem[{{Pols} \& {Dewi}(2002)}]{pols02}
{Pols}, O.~R. \& {Dewi}, J.~D.~M. 2002, Publications of the Astronomical
  Society of Australia, 19, 233

\bibitem[{{Portegies Zwart} \& {Verbunt}(1996)}]{zwart96}
{Portegies Zwart}, S.~F. \& {Verbunt}, F. 1996, \aap, 309, 179

\bibitem[{{Rappaport} {et~al.}(1983){Rappaport}, {Joss}, \&
  {Verbunt}}]{rappaport83b}
{Rappaport}, S., {Joss}, P.~C., \& {Verbunt}, F. 1983, \apj, 275, 713

\bibitem[{{Rappaport} {et~al.}(1982){Rappaport}, {Joss}, \&
  {Webbink}}]{rappaport82b}
{Rappaport}, S., {Joss}, P.~C., \& {Webbink}, R.~F. 1982, \apj, 254, 616

\bibitem[{{Ruderman} {et~al.}(1989){Ruderman}, {Shaham}, \&
  {Tavani}}]{ruderman89}
{Ruderman}, M., {Shaham}, J., \& {Tavani}, M. 1989, \apj, 336, 507

\bibitem[{{Scalo}(1986)}]{scalo86a}
{Scalo}, J.~M. 1986, Fundamentals of Cosmic Physics, 11, 1

\bibitem[{{Shirey}(1998)}]{shirey98}
{Shirey}, R.~E. 1998, Ph.D.~Thesis, MIT

\bibitem[{{Taam} {et~al.}(2000){Taam}, {King}, \& {Ritter}}]{taam00}
{Taam}, R.~E., {King}, A.~R., \& {Ritter}, H. 2000, \apj, 541, 329

\bibitem[{{Tauris} \& {Savonije}(1999)}]{tauris99}
{Tauris}, T.~M. \& {Savonije}, G.~J. 1999, \aap, 350, 928

\bibitem[{{Tauris} {et~al.}(2000){Tauris}, {van den Heuvel}, \&
  {Savonije}}]{tauris00}
{Tauris}, T.~M., {van den Heuvel}, E.~P.~J., \& {Savonije}, G.~J. 2000, \apjl,
  530, L93

\bibitem[{{Thorsett} \& {Chakrabarty}(1999)}]{thorsett99}
{Thorsett}, S.~E. \& {Chakrabarty}, D. 1999, \apj, 512, 288

\bibitem[{{van der Klis}(2000)}]{vanderklis00}
{van der Klis}, M. 2000, \araa, 38, 717

\bibitem[{{van Paradijs}(1996)}]{vanparadijs96}
{van Paradijs}, J. 1996, \apjl, 464, L139+

\bibitem[{{Verbunt} \& {Zwaan}(1981)}]{verbunt81}
{Verbunt}, F. \& {Zwaan}, C. 1981, \aap, 100, L7

\bibitem[{{Vrtilek} {et~al.}(1990){Vrtilek}, {Raymond}, {Garcia}, {Verbunt},
  {Hasinger}, \& {Kurster}}]{vrtilek90}
{Vrtilek}, S.~D., {Raymond}, J.~C., {Garcia}, M.~R., {Verbunt}, F., {Hasinger},
  G., \& {Kurster}, M. 1990, \aap, 235, 162

\bibitem[{{Webbink}(1984)}]{webbink84}
{Webbink}, R.~F. 1984, \apj, 277, 355

\bibitem[{{Wellstein} {et~al.}(2001){Wellstein}, {Langer}, \&
  {Braun}}]{wellstein01}
{Wellstein}, S., {Langer}, N., \& {Braun}, H. 2001, \aap, 369, 939

\bibitem[{{Wijnands} \& {van der Klis}(1998)}]{wijnands98}
{Wijnands}, R. \& {van der Klis}, M. 1998, \nat, 394, 344

\bibitem[{{Willems} \& {Kolb}(2002)}]{willems02}
{Willems}, B. \& {Kolb}, U. 2002, \mnras, 337, 1004

\end{thebibliography}


\end{document}